\documentclass[aps,prd,onecolumn,groupedaddress,showpacs,nofootinbib,amssymb
]{revtex4}
\usepackage[dvips]{graphicx}
\usepackage{amssymb}
\usepackage{amsmath}
\usepackage{graphicx}
\usepackage{amsfonts}
\usepackage{bm}

\begin{document}

\title{Deformed Matter Bounce with Dark Energy Epoch}
\author{
S.~D.~Odintsov,$^{1,2}$\,\thanks{odintsov@ieec.uab.es}
V.~K.~Oikonomou,$^{3,4}$\,\thanks{v.k.oikonomou1979@gmail.com}}
\affiliation{ $^{1)}$Institut de Ciencies de lEspai (IEEC-CSIC),
Campus UAB, Carrer de Can Magrans, s/n\\
08193 Cerdanyola del Valles, Barcelona, Spain\\
$^{2)}$ ICREA, Passeig LluA­s Companys, 23,
08010 Barcelona, Spain\\
$^{3)}$ Tomsk State Pedagogical University, 634061 Tomsk, Russia\\
$^{4)}$ Laboratory for Theoretical Cosmology, Tomsk State University of Control Systems
and Radioelectronics (TUSUR), 634050 Tomsk, Russia\\
}

\begin{abstract}
We extend the Loop Quantum Cosmology matter bounce scenario in order to include a dark energy era, which ends abruptly at a Rip singularity where the scale factor and the Hubble rate diverge. In the ``deformed matter bounce scenario'', the Universe is contracting from an initial non-causal matter dominated era until it reaches a minimal radius. After that it expands in a decelerating way, until at late times, where it expands in an accelerating way, thus the model is described by a dark energy era that follows the matter dominated era. Depending on the choice of the free parameters of the model, the dark energy era is quintessential like which follows the matter domination era, and eventually it crosses the phantom divide line and becomes phantom. At the end of the dark energy era, a Rip singularity exists, where the scale factor and Hubble rate diverge, however the physical system cannot reach the singularity, since the effective energy density and pressure become complex. This indicates two things, firstly that the ordinary Loop Quantum Cosmology matter bounce evolution stops, thus ending the infinite repetition of the ordinary matter bounce scenario. Secondly, the fact that both the pressure and the density become complex indicates probably that the description of the cosmic evolution within the theoretical context of Loop Quantum Cosmology, ceases to describe the physics of the system and possibly a more fundamental theory of quantum gravity is needed near the would be Rip singularity. We describe the qualitative features of the model and we also investigate how this cosmology could be realized by a viscous fluid in the context of Loop Quantum Cosmology. In addition to this, we show how this deformed model can be realized by a canonical scalar field filled Universe, in the context of Loop Quantum Cosmology. Finally, we demonstrate how the model can be generated by a vacuum $F(R)$ gravity. 
\end{abstract}


\maketitle



\def\pp{{\, \mid \hskip -1.5mm =}}
\def\cL{\mathcal{L}}
\def\be{\begin{equation}}
\def\ee{\end{equation}}
\def\bea{\begin{eqnarray}}
\def\eea{\end{eqnarray}}
\def\tr{\mathrm{tr}\, }
\def\nn{\nonumber \\}
\def\e{\mathrm{e}}

\section{Introduction}

The inflationary paradigm for early-time cosmology is a consistent framework, in which many theoretical inconsistencies of the Big Bang cosmology were appropriately addressed \cite{inflation1,inflation2,inflation3,inflation4,inflation5}. However, inflation theories suffer from the initial singularity problem \cite{vilenkin} at the origin $t=0$, and this initial singularity is a crushing type timelike singularity \cite{hawkingpenrose}, which means that the spacetime is geodesically incomplete at that time instance. An appealing alternative scenario to the standard inflationary cosmology, is the big bounce scenario, in the context of which the singularity is replaced by a bounce \cite{brande1,bounce1,bounce1a,bounce1b,bounce2,bounce3,bounce4,bounce5,bounce6,bounce7,bounce8,bounce9,bounce10,bounce11,bounce12,bounce13,bounce14,bounce16,matterbounce1,matterbounce2,matterbounce4,matterbounce5,matterbounce6,matterbounce7,matterbounce8}. Particularly, in most bouncing cosmologies, the Universe initially is large and cold and starts its evolution contracting until it reaches a minimal radius, at which point it bounces off and starts to expand again. Among the most interesting bouncing scenarios is the matter bounce cosmological scenario \cite{matterbounce1,matterbounce2,matterbounce4,matterbounce5,matterbounce6,matterbounce7,matterbounce8}, in the context of which, the Universe starts contracting and is described by a matter dominated initial cosmological era and the spectrum is scale invariant \cite{brande1}. Initially, the Hubble horizon is large and thus the perturbations are at sub-Hubble scales, and as the Universe contracts, these modes exit the horizon and thus will become relevant for present day observations after these reenter the horizon during the expanding phase, after the bouncing point at $t=0$. The contraction phase is very important for the phenomenological properties of the Universe, since during this phase, the various parts of the Universe come to a causal contact. However, the matter bounce cosmologies, like most bouncing cosmologies, have certain pathologies which should be appropriately addressed in a consistent theoretical way in order the theory becomes complete. Examples of these pathologies are the fact that the spectrum is exactly scale invariant, the primordial non-gaussianities, the suppression of the primordial gravitational waves quantified in terms of the scalar-to-tensor ratio and the late-time acceleration era, see \cite{brande1} for a detailed presentation on these issues. 

The purpose of this paper is to address the late-time acceleration issue of the matter bounce scenario, in the context of Loop Quantum Cosmology (LQC) \cite{LQC1,LQC3,LQC4,LQC5,LQC6,LQC7,LQC9,LQC10,LQC11,LQC12,LQC13,LQC14,LQC15}, and also in modified gravity. Particularly, we shall deform the matter bounce scenario in a way so that the late-time acceleration occurs in the model, and we shall refer to this model as the deformed matter bounce scenario. According to the deformed matter bounce scenario, the Universe starts the contracting phase at $t\rightarrow -\infty $ while being described by the matter bounce scenario, since the deformation effects are negligible at the beginning of the contraction, and then it bounces off at $t=0$ and starts to expand again. Gradually, as the Universe expands, it is described by a matter dominated epoch, and the deformation of the matter bounce scenario affects the cosmological evolution only at late times, where a dark energy era occurs. The dark energy era is quintessential and the Universe crosses the phantom divide at late times, so eventually the dark energy era is phantom. We need to note that the latest observations coming from Planck data \cite{planckphant}, but also earlier data coming from WMAP \cite{wmap} indicate that the total equation of state parameter for dark energy is $w_{eff}=-1.54^{+0.62}_{-0.50}$, which is almost $2\sigma$ in the phantom domain \cite{planckphant}.

At the end of the phantom dark energy era, at a finite time $t_s$, the scale factor and the Hubble rate of the model strongly diverge and this indicates that a Big Rip or some sort of Rip singularity \cite{ref51,ref52,ref53,ref54,ref55,ref56,ref57,Nojiri:2005sx} is reached. However, a few seconds before this Rip singularity, the effective energy density and the effective pressure of the model become complex and thus this singularity is unphysical. We conclude that the Rip singularity is not accessible for the physical system, and therefore this behavior of the very late-time era clearly indicates that even in the LQC context this era is effectively described by a more fundamental theory of quantum gravity. This behavior is reminiscent to results studied in related works \cite{nobigrip1,nobigrip2,nobigrip3,nobigrip4,haronobigrip}, which state that the Rip singularity is avoided in LQC. In our case the Rip is not avoided directly, but the physical system cannot reach this singularity, since some observables become complex and thus unphysical. At a later point we shall discuss this issue in detail since it requires a careful interpretation.

The motivation for studying singular deformations of the matter bounce scenario comes mainly from the fact that recent constraints hint that the total equation of state parameter is around $w_{eff}\simeq -1$ (slightly larger or smaller than $w_{eff}=-1$) and also from the fact our Universe undergoes a late-time acceleration era. The standard matter bounce scenario does not have these features, so with this paper we aim to demonstrate that this can happen by deforming the standard matter bounce scenario. As we show in the next sections, the deformed matter bounce scenario and the standard matter bounce scenario differ only at late times.

This paper is organized as follows: In section II, we present in some detail all the qualitative new features of the deformed matter bounce cosmology and we discuss the behavior of the EoS and also we describe the late-time acceleration era. We also investigate the phase space of our model and also we examine which is the final attractor of our model. The quantitative analysis is supported by numerical results too, in order to further understand the resulting qualitative features of the model. In section III, we investigate which LQC viscous imperfect fluid can realize the deformed matter bounce cosmology and we do the same in the case of a canonical scalar field in the context of LQC. In section IV, using known reconstruction techniques, we shall investigate how it is possible to realize the deformed matter bounce cosmology in the context of vacuum $F(R)$ gravity, emphasizing in the description of the late-time era, since the early-time description already exists in the literature.

Before getting started, let us briefly present here the geometric conventions we shall assume for the background spacetime. We will use a flat Friedmann-Robertson-Walker background, with line element,
\be
\label{metricfrw} ds^2 = - dt^2 + a(t)^2 \sum_{i=1,2,3}
\left(dx^i\right)^2\, ,
\ee
where as usual, $a(t)$ denotes the scale factor. In addition, we shall consider that the connection is a symmetric, torsion-less and metric compatible affine connection, the Levi-Civita connection. For the metric (\ref{metricfrw}), the Ricci scalar reads,
\begin{equation}
\label{ricciscal}
R=6(2H^2+\dot{H})\, ,
\end{equation}
with $H$ being the Hubble rate $H=\dot{a}/a$. Also we use a units system such that $\hbar=c=8\pi G=\kappa^2=1$.

\section{The Qualitative Features of the Deformed Matter Bounce Scenario}

The LQC framework \cite{LQC1,LQC3,LQC4,LQC5,LQC6,LQC7,LQC9,LQC10,LQC11,LQC12,LQC13,LQC14,LQC15} offers a theoretical description of cosmological evolution in which the initial singularity is avoided. In the context of LQC, the effective Hamiltonian that describes the quantum Universe is \cite{LQC1,LQC3,LQC4,LQC5,LQC6,LQC7,LQC9,LQC10,LQC11,LQC12,LQC13,LQC14,LQC15},
\begin{equation}\label{effhamilt}
\mathcal{H}_{LQC}=-3V\frac{\sin^2(\lambda \beta)}{\gamma^2\lambda^2}+V\rho\, ,
\end{equation}
with $\gamma$ being the Barbero-Immirzi parameter, $\lambda$ a parameter related to $\gamma$, with dimensions of length, $V$ is the volume $V=a(t)^3$, $a(t)$ the scale factor, and $\rho$ the energy density of the matter which fills the Universe. By using the Hamiltonian constraint $\mathcal{H}_{LQC}=0$, that is,
\begin{equation}\label{hamiltonianconstr}
\frac{\sin^2(\lambda \beta)}{\gamma^2\lambda^2}=\frac{\rho}{3}\, ,
\end{equation}
and also using the following anti-commutation identity,
\begin{equation}\label{anticom}
\dot{V}=\{V,\mathcal{H}_{LQC}\}=-\frac{\gamma}{2}\frac{\partial \mathcal{H}_{LQC}}{\partial \beta}, 
\end{equation}
we obtain the holonomy quantum corrected Friedmann equation \cite{LQC1,LQC3,LQC4,LQC5,LQC6,LQC7,LQC9,LQC10,LQC11,LQC12,LQC13,LQC14,LQC15},
\begin{equation}\label{holcor1}
H^2=\frac{\rho}{3}\left (1-\frac{\rho}{\rho_c}\right )\, ,
\end{equation} 
where the energy density $\rho$ is assumed to satisfy the continuity equation,
\begin{equation}\label{cont}
\dot{\rho}(t)=-3H\Big{(}\rho(t)+P(t) \Big{)}\, ,
\end{equation}
and with $P(t)$ being the effective pressure of the perfect matter fluid. By using the continuity equation and the holonomy corrected Friedman equation, we obtain the following differential equation too,
\begin{equation}\label{eqnm}
\dot{H}=-\frac{1}{2}(\rho+P)(1-2\frac{\rho}{\rho_c})\, .
\end{equation}
Notice that both Eqs. (\ref{holcor1}) and (\ref{eqnm}), in the limit $\rho_c\to \infty$, lead to the ordinary Friedmann equations of Einstein-Hilbert gravity, so the finiteness of the parameter $\rho_c$ captures the quantum nature of spacetime. The appealing scenario which we shall be interested is the matter bounce scenario \cite{matterbounce1,matterbounce2,matterbounce4,matterbounce5,matterbounce6,matterbounce7,matterbounce8}, see also \cite{lcdmcai,lcdmsergei} for a presentation of the alternative radiation bounce scenario in LQC. In the case of the matter bounce scenario, the pressure is zero, so the effective energy density $\rho_{mb}(t)$ satisfies,
\begin{equation}\label{cont1}
\dot{\rho}_{mb}(t)=-3H\rho_{mb}(t)\, .
\end{equation}
The differential equation (\ref{cont1}) has the well known solution $\rho_{mb}=\rho_c a^{-3}$, which describes a matter dominated Universe. By plugging the expression $\rho_{mb}=\rho_c a^{-3}$ in the LQC-corrected Friedmann equation, the scale factor and the corresponding Hubble rate can easily be found and these read,
\begin{equation}\label{holcorrLQCsol}
a_{mb}(t)=\left (\frac{3}{4}\rho_ct^2+1\right )^{1/3},{\,}{\,}{\,}H_{mb}(t)=\frac{\frac{1}{2}\rho_ct}{\frac{3}{4}\rho_ct^2+1}\, .
\end{equation}
By using the resulting form of the scale factor from Eq. (\ref{holcorrLQCsol}) and plugging it in the expression $\rho_{mb}=\rho_c a^{-3}$, the matter energy density as a function of the cosmic time becomes, 
\begin{equation}\label{rhosol}
\rho_{mb} (t)=\frac{\rho_c}{\frac{3}{4}\rho_c t^2+1}\, .
\end{equation}
In this work we propose a deformed version of the matter bounce scenario, in which case the deformed matter bounce scale factor $a(t)$ and in effect, the corresponding Hubble rate $H(t)$, have the following form,
\begin{equation}\label{deformedscalefactor}
a(t)=\left (\frac{3}{4}\rho_ct^2+1\right )^{1/3}\times e^{\frac{f_0}{1-\alpha}\left(t-t_s\right)^{1-\alpha}},\,\,\, H(t)=\frac{\frac{1}{2}\rho_ct}{\frac{3}{4}\rho_ct^2+1}+f_0\left(t-t_s \right)^{-\alpha}\, ,
\end{equation}
where $t_s$ is a time instance, the value of which we now discuss, since this will play some important role in the analysis. From the form of the deformed Hubble rate and scale factor in Eq. (\ref{deformedscalefactor}), it can be seen that the values of the parameter $\alpha$ crucially determine the finite time singularity structure of the theory. Particularly, according to the values of $\alpha$, if $\alpha>1$, we now demonstrate that a Type I \cite{Nojiri:2005sx}, or so called Big Rip singularity can occur at $t_s$, since both the Hubble rate and the scale factor diverge at $t=t_s$. Note however that for the Big Rip singularity to occur, it is required that both the energy density and the pressure also diverge at that time instance. As we shall see, the physical picture in the context of LQC is complicated as the time instance $t_s$ is approached, so we need to carefully present the resulting picture. As we show, the singularity is a Rip for sure, but what sort of Rip cannot be easily determined.

Before we proceed with the singularity type, we need to discuss the important issue of choosing the values of $\alpha$ correctly. We will be interested in the case of a Big Rip singularity at $t=t_s$, therefore $\alpha>1$ and also it will considered to have the form $\alpha=2n/(2m+1)$, where $m,n$ positive integers, in order to avoid complex values in the Hubble rate. Also we assume that the time instance $t_s$ corresponds to a late-time era and particularly, that the Big Rip singularity occurs at the very late times, for $t\rightarrow t_s$, which could be some time instance far beyond the present time era ($13.6$ billion years), for example $t_s\simeq 10^{18}$sec, which corresponds to a Big Rip time instance that occurs when the Universe is approximately $32.5$ billion years old. This point requires a clarification in order to avoid unnecessary complications, with regards to the behavior of the scale factor and also with regards to the values of $\alpha$. When $\alpha$ is chosen to be $\alpha=2n/(2m+1)$, then the exponential in the scale factor becomes of the form $e^{\frac{f_0}{1-\alpha}\left(t-t_s\right)^{\frac{2(m-n)+1}{2n+1}}}$, so since $t<t_s$, this means that the exponential behaves as $e^{-\frac{f_0}{1-\alpha}\left(t_s-t\right)^{\frac{2(m-n)+1}{2n+1}}}$, since the number $\frac{2(m-n)+1}{2n+1}$ consists of odd numbers\footnote{Recall that $(t-t_s)^{\frac{2(m-n)+1}{2n+1}}$ is not complex, since for example $(-1)^{5/3}$ has two complex conjugate roots and one real but negative. This explains the negative sign of the exponential.} Therefore, complex values of the scale factor are avoided with this choice of $\alpha$. 

Having specified the allowed values for the parameter $\alpha$, we now demonstrate that the singularity at $t=t_s$ is indeed a Big Rip, or at least it is similar to a Big Rip, where the scale factor and Hubble rate diverge. Indeed, by using the LQC first Friedmann equation, namely Eq. (\ref{holcor1}), since the Hubble rate in Eq. (\ref{deformedscalefactor}) strongly diverges at $t=t_s$ for $\alpha>1$, this means that the total energy density $\rho$ also diverges. In addition, from the second Friedmann equation, namely Eq. (\ref{eqnm}), the total pressure density is equal to,
\begin{equation}\label{metviw1}
P=-\rho-\frac{2\dot{H}}{1-\frac{2\rho}{\rho_c}}\, ,
\end{equation}
and since $\dot{H}$ and also $\rho$ diverge at $t=t_s$ for the Hubble rate being (\ref{deformedscalefactor}), this means that also the pressure diverges at $t=t_s$. So at $t=t_s$, the scale factor, the total energy density and total pressure diverge and according to the classification made in Ref. \cite{Nojiri:2005sx}, this finite time singularity corresponds to a Big Rip, or Type I singularity. Actually, in the list below, we quote the singularity structure of the cosmological evolution (\ref{deformedscalefactor}), in the context of LQC theory governed by the Friedman equations (\ref{holcor1}) and (\ref{eqnm}).
\begin{itemize}\label{lista}
\item For $\alpha<-1$, a Type IV singularity occurs at $t_s$.
\item For $-1<\alpha<0$, a Type II singularity occurs at $t_s$
\item For $0<\alpha<1$, a Type III singularity occurs at $t_s$
\item For $\alpha>1$, a Type I, or so called Big Rip singularity occurs at $t_s$. 
\end{itemize}
However, at the actual time instance $t=t_s$, both the energy density and pressure do not only diverge, but also acquire complex values. This indicates that the actual singularity cannot be reached from our physical system, since there is possibly another theory, more fundamental than the LQC, that governs this era. As we will demonstrate, the era for which the complex behavior of the energy and pressure occurs starts one second before the actual singularity is reached, hence limiting the applicability of the LQC theoretical description after that time instance. Note however that even in classical descriptions \cite{kamio}, the Universe experiences the catastrophic effects of a Big Rip or similar Rip singularity, long before the singularity is reached, for example a million years before the Rip, the clusters are stripped, and therefore nearly a second before the Rip, the Universe is in a plasma state, and possibly a more fundamental quantum gravity description is needed for the description of the physical picture. 

In our case, the fact that the energy and density become complex very close to the Rip singularity clearly indicates that, firstly the Rip is not reached. Secondly, the physical description of the model is governed by another more fundamental theory than LQC, which should govern the physics approximately one second before the Rip, as numerical analysis of our model shows. The important feature of the deformed matter bounce model we propose is that the dark energy era, is consistently described and follows the matter dominated era. Also of equal importance is that the matter bounce scenario is interrupted one second before the Rip, so the infinite repetition of a matter bounce Universe is avoided. Also the viewpoint that the Rip is not reached by the physical system, agrees with other LQC theoretical descriptions that state exactly this, that the Rip singularity does not occur. In our case, the difference is that the LQC description ceases to be valid very close to the Rip, so as a side effect, the Rip singularity is avoided, at least in the context of the LQC description.

Before proceeding it is worth discussing another interesting possibility for the cosmological evolution (\ref{deformedscalefactor}), namely the possibility of a big crush or little rip or even a Pseudo-Rip \cite{newrips1,newrips2,newrips3}. It seems that the singularity is a Big Rip, but this is a delicate issue as we discuss shortly. The inertial force $F_{in}$ between a comoving observer and a mass, which is at a distance $L$ from the observer is equal to,
\begin{equation}\label{inertialforcecomov}
F_{in}=mL\frac{\ddot{a}}{a}=mL\left(\dot{H}+H^2 \right)\, ,
\end{equation}
and if either $H$ or $\dot{H}$ diverge, a Rip always occurs. In the case that both $H$ and $\dot{H}$ diverge, with $\dot{H}>0$, this corresponds to the Big Rip case, which seems to be our case. In the case that $H$ is not singular at finite time, but $\dot{H}$ is singular, with $\dot{H}>0$, then this corresponds to a sudden future singularity or Type II \cite{sergeirip}. For the cosmology (\ref{deformedscalefactor}), the analytic form of $\dot{H}$ at late times before $t=t_s$, is approximately equal to,
\begin{equation}\label{dothlatetime}
\dot{H}\simeq -\frac{2}{3 t^2}-f_0 (t-t_s)^{-1-\alpha } \alpha\, ,
\end{equation}
and since $\alpha=2n/(2m+1)$, the sign of $\dot{H}$ is approximately equal to,
\begin{equation}\label{dothlatetime1}
\dot{H}\simeq -\frac{2}{3 t^2}+f_0 (t_s-t)^{-\frac{2(m+n)+1}{2m+1} } \alpha\, ,
\end{equation}
and since the dominant contribution comes from the second term, then $\dot{H}\simeq f_0 (t_s-t)^{-\frac{2(m+n)+1}{2m+1} } \alpha$, which is clearly positive for $t<t_s$. Therefore, the inertial force diverges at finite time with $\dot{H}>0$, which corresponds to a Big Rip behavior, as we anticipated. In addition, there is no room for crushing type phenomena for our cosmological scenario, since $\dot{H}>0$ before the Big Rip singularity, since in the case of crushing phenomena we would have $\dot{H}<0$. Note also that in the case of Little Rip, $F_{in}\to \infty$, which also occurs in the case of a Big Rip, with the difference that in the Little Rip, the infinite inertial force is obtained in the limit $t\to \infty$, whereas in the Big Rip case, the inertial force tends to infinity at a finite time, which is the case at hand. This is also a strong indication that a Big Rip behavior occurs in the future of our model, however the physical system cannot experience it at the time it occurs, nevertheless everything is pointing out to a late-time Big Rip.

An important comment is in order. Formally we have asymptotically de Sitter limit where the Universe enters this era being always in a phantom state. This is well-known as the Little Rip case. However, this limit is just mathematical, it is impossible to reach the de Sitter point in reality, due to the following. In physical theories with Pseudo-Rip, Big Rip or Little Rip late-time behavior, always occurs the dissolution of bound objects well before the singularity occurs. For our case then, regardless if the singularity is a Big Rip or other, before the singularity is approached, the repulsive nature of the dark energy will destroy every bound object well before the singularity.

We now proceed to the analysis of the deformed matter bounce model, and for convenience and notational simplicity, we write the deformed scale factor and Hubble rate in the following form,
\begin{equation}\label{deformconven}
a(t)=a_{mb}(t)\times a_s(t),\,\,\,H(t)=H_{mb}(t)+H_s(t)\, ,
\end{equation}
where $a_{mb}(t)$ and $H_{mb}(t)$ are defined in Eq. (\ref{holcorrLQCsol}), and $a_s(t)$ and $H_s(t)$ are equal to,
\begin{equation}\label{ashs}
a_s(t)=e^{\frac{f_0}{1-\alpha}\left(t-t_s\right)^{1-\alpha}},\,\,\, H_s(t)=f_0\left(t-t_s \right)^{-\alpha}\, ,
\end{equation}
and these characterize the singular deformed parts of the Hubble rate and scale factor. For the values we constrain $\alpha$ to take, the cosmology of Eq. (\ref{deformedscalefactor}) clearly describes a bounce, since the function $H_s(t)$ does not affect the behavior of the Hubble rate until late times, and the same applies for the scale factor. 
\begin{figure}[h] \centering
\includegraphics[width=15pc]{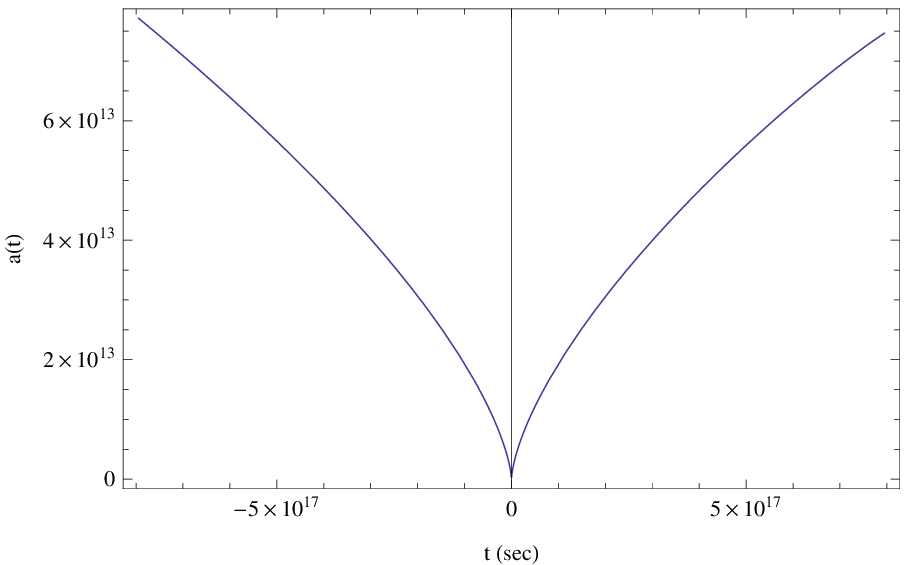}
\includegraphics[width=15pc]{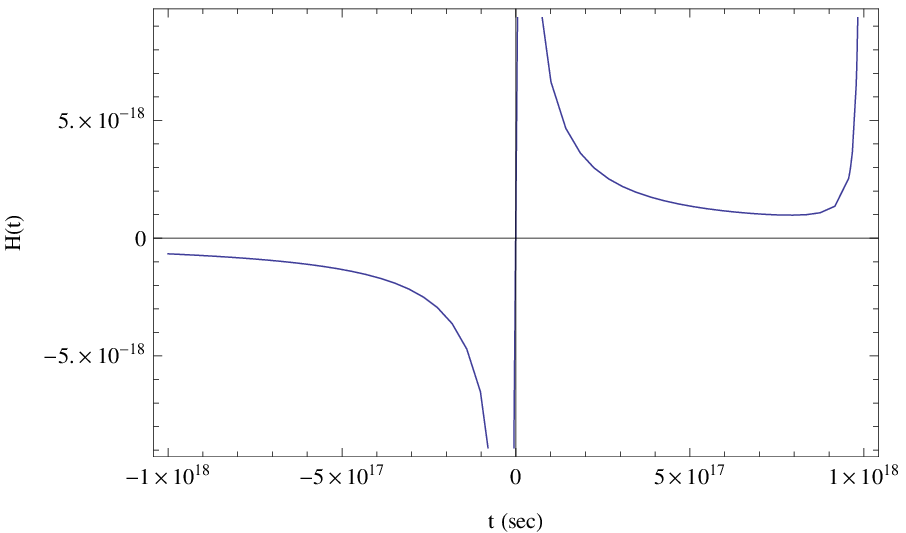}
\caption{The Hubble rate $H(t)$ (left) and the scale factor $a(t)$ (right) versus time, for $\alpha=4/3$, $t_s=10^{18}$sec and $\rho_c/f_0\simeq 10^{-4}$. Notice that at the Rip singularity the cosmological evolutions is abruptly interrupted.}
\label{plota}
\end{figure}
This can also be seen in Fig. \ref{plota}, where we plotted the Hubble rate $H(t)$ and the scale factor $a(t)$ as functions of time, for $\alpha=4/3$, and $f_0$ and $\rho_c$ chosen is such way so that $\rho_c/f_0\simeq 10^{-4}$, and from the two plots it can be seen that the bouncing behavior is not altered and persists, without being affected from the singular deformation part, at least until late times at $t\simeq 10^{18}$sec, where the cosmological evolution is abruptly interrupted. 
\begin{figure}[h] \centering
\includegraphics[width=15pc]{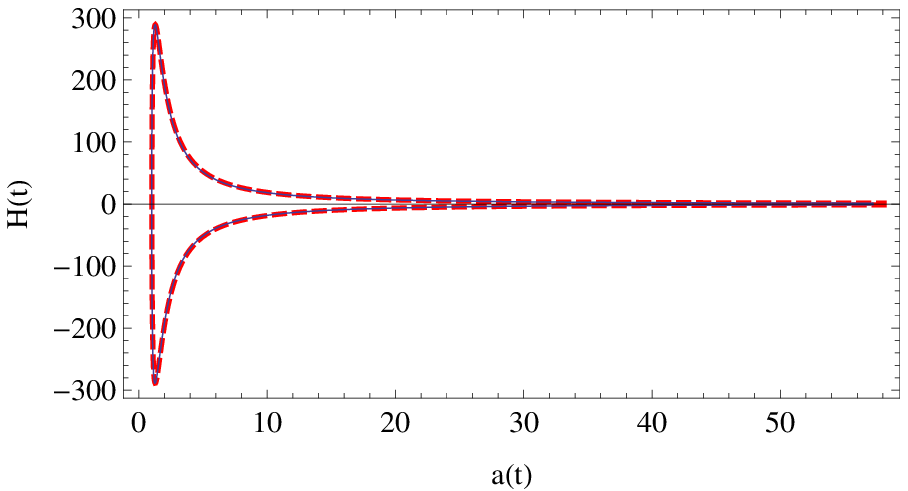}
\includegraphics[width=15pc]{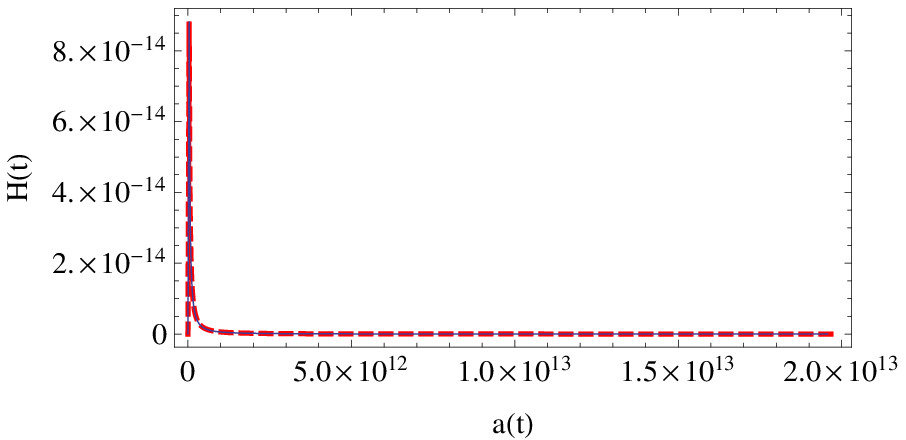}
\includegraphics[width=15pc]{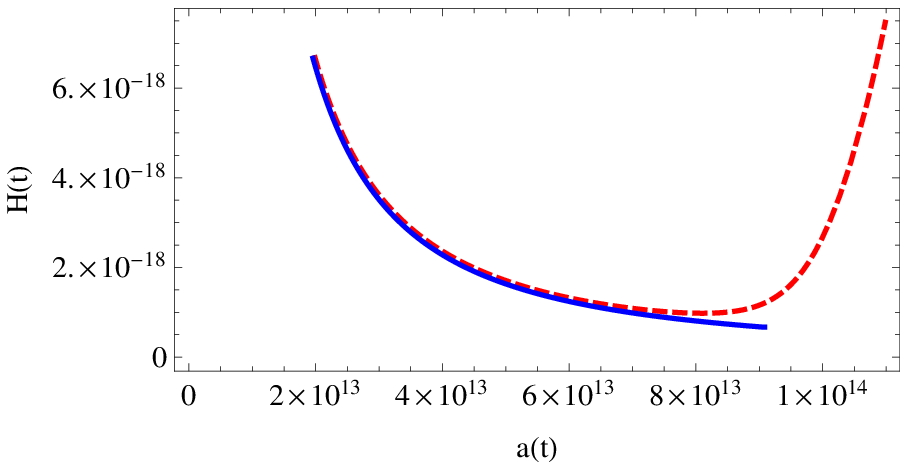}
\caption{Comparison of the parametric plots (left up plot) of the scale factor $a(t)$ versus the Hubble rate $H(t)$, for the deformed matter bounce (red, dashed curve) and the ordinary matter bounce scenario (blue curve). We have used the values $\alpha=4/3$, $t_s=10^{18}$sec and $\rho_c/f_0\simeq 10^{-4}$ and the parametric plot corresponds to the time interval $0<t<10$ sec. In the upper right and left plots appears the parametric plot for the deformed matter bounce scenario and in the bottom plot appears the ordinary matter bounce case (blue curve) and the deformed matter bounce (red curve) for the time interval this time is $10^{17}<t<9.8\times 10^{17}$ sec.}
\label{plotrev1}
\end{figure}
Instead of the plots appearing in Fig. \ref{plota} it would be more useful to plot directly the Hubble rate as a function of the scale factor. However in order to find the explicit functional dependence of the Hubble rate as a function of the scale factor we need to find the function $t=t(a)$ by inverting the scale factor in Eq. (\ref{deformedscalefactor}) and then substitute the cosmic time in the expression for the Hubble rate appearing in Eq. (\ref{deformedscalefactor}). This task is however formidable since it is not easy to invert the function $a(t)$, for a general $\alpha$ (it can be done for $\alpha=1$, which is not an interesting case since we need a Big Rip singularity). Therefore in order to have a clear picture of the $H(a)$ behavior we perform the parametric plot $a(t)-H(t)$, which can be found in the left plot of Fig. \ref{plotrev1}.  The blue curve corresponds to the ordinary matter bounce scenario while the red dashed curve corresponds to the deformed matter bounce. We have used values $\alpha=4/3$, $t_s=10^{18}$sec and $\rho_c/f_0\simeq 10^{-4}$ and the parametric plot corresponds to the time interval $0<t<10$ sec, but the qualitative behavior remains the same up to approximately $t=10^{17}$sec.

As it can be seen in the left and right plots of Fig. \ref{plotrev1}, the parametric plot of $a(t)-H(t)$ for the deformed bounce and the matter bounce remains the same until the late times, where the deformed matter bounce changes drastically the resulting picture. The left plot corresponds to the time interval $-10<t<10$sec, while the right to the interval $0<t<10^{17}$sec. Notice that the left plot has the negative Hubble rate branch which corresponds to the contracting phase of the bounce. The right plot contains only the expanding phase, so the behavior of both the matter bounce and the deformed matter bounce model remains the same up to $t=10^{17}$sec. After that time instance, the deformed matter bounce starts to deviate from the ordinary matter bounce scenario, and this can be seen in the bottom plot of Fig. \ref{plotrev1}, which corresponds to the time interval $10^{17}<t<9.8\times 10^{17}$ sec. As it can be seen in the bottom plot of  Fig. \ref{plotrev1}, the red curve starts to deviate from the blue curve (ordinary matter bounce), and this happens approximately at $t=9.5\times 10^{17}$sec. We need to note that the Hubble rate in the deformed matter bounce scenario increases after some point, since a Big Rip singularity is approached, so the Hubble rate will blow up at the singularity. So as the singularity is approached, the Hubble rate increases, as the scale factor also increases. We shall come back to this issue in a later section, where we shall study the behavior of the expression $R_H=1/(a(t)H(t))$ as a function of the cosmic time.

With the assumption that a Rip singularity occurs at the very late times, when $t\rightarrow t_s$, the following relations holds true, as it can be easily seen from Eq. (\ref{deformedscalefactor}),
\begin{align}\label{limits}
\lim_{t\to -\infty}H_s(t)\simeq 0,\,\,\,\lim_{t\to -\infty}a_s(t)\simeq 1,\,\,\,\lim_{t\to -\infty}H(t)\simeq H_{mb}(t),\,\,\,\lim_{t\to t_s}a(t)\simeq a_s(t),\,\,\,\lim_{t\to t_s}H(t)\simeq H_{s}(t)\, ,
\end{align}
and the Hubble rate and scale factor blow up as $t\to t_s$. Effectively, the singular deformation part of the scale factor and of the Hubble rate in Eq. (\ref{deformedscalefactor}) affect the cosmological evolution only at very late-times, and at earlier times the matter bounce scenario occurs, without being seriously affected by the singular deformation part. Notice that the assumption $t_s\gg 1$ is crucial for obtaining the limits of Eq. (\ref{limits}), since the exponential term in the scale factor is of the order $e^{t_s^{\alpha}}\simeq \mathcal{O}(1)$, and this holds true until late times, where $t\simeq t_s$. The next step is to find which matter-energy density can produce the cosmological evolution (\ref{deformedscalefactor}), in the context of LQC. We denote the deformed energy density as $\rho(t)$ and we assume that it has the form $\rho(t)=\rho_{mb}(t)+\rho_s(t)$, where $\rho_{mb}$ is defined in Eq. (\ref{rhosol}) and it satisfies the continuity equation (\ref{cont1}). The effective pressure of the deformed matter fluid is denoted by $P$ so it satisfies the continuity equation,
\begin{equation}\label{cont2}
\dot{\rho}=-3H(\rho+P)\, ,
\end{equation}
and also it satisfies the holonomy corrected Friedmann equation (\ref{holcor1}). Therefore, by combining Eqs. (\ref{holcor1}), (\ref{cont1}) and (\ref{cont2}), we can easily read off the energy density and the effective pressure of the deformed matter fluid,
\begin{equation}\label{defomedmatt}
\rho_s(t)=-\rho_{mb}(t)+\frac{\rho_c-\sqrt{\rho_c^2-12\rho_cH(t)^2}}{2},\,\,\,P(t)=-\frac{\dot{\rho}_s(t)+3H_s(t)\rho_{mb}(t)+3H(t)\rho_s(t)}{3H}\, .
\end{equation} 
The singular deformed energy density and the corresponding effective pressure, at $t\to -\infty $, tend to describe the matter bounce perfect fluid, as it can be seen by using the limits of Eq. (\ref{limits}). Hence, with the effective fluid (\ref{defomedmatt}) we achieve two important things: firstly we have a cosmological evolution which is described by the matter bounce scenario for all eras, except for the late-time era, and secondly we include a Rip singularity occurring at late times, in the context of a non-singular bouncing cosmology at the origin $t=0$. With our description therefore we have all the good features of the matter bounce scenario, that is, a scale invariant spectrum at early times, the avoidance of the initial singularity at $t=0$, and the new feature, the end of the infinite bouncing cycle at the late-time Rip singularity. With regards to the power spectrum during the contracting phase and the proof that it is scale invariant, we refer the reader to the recent review \cite{brande1} for a detailed analysis of this issue. Note however that the energy density (\ref{defomedmatt}) becomes complex at the vicinity of the Rip singularity and therefore the theory fails to describe the vicinity of the Rip. This is why we claimed that the matter bounce scenario description stops, since another more fundamental theory is needed to describe the physics at the vicinity of the Rip singularity. In any case, the infinite repetition of the matter bounce scenario is avoided.

\begin{figure}[h] \centering
\includegraphics[width=15pc]{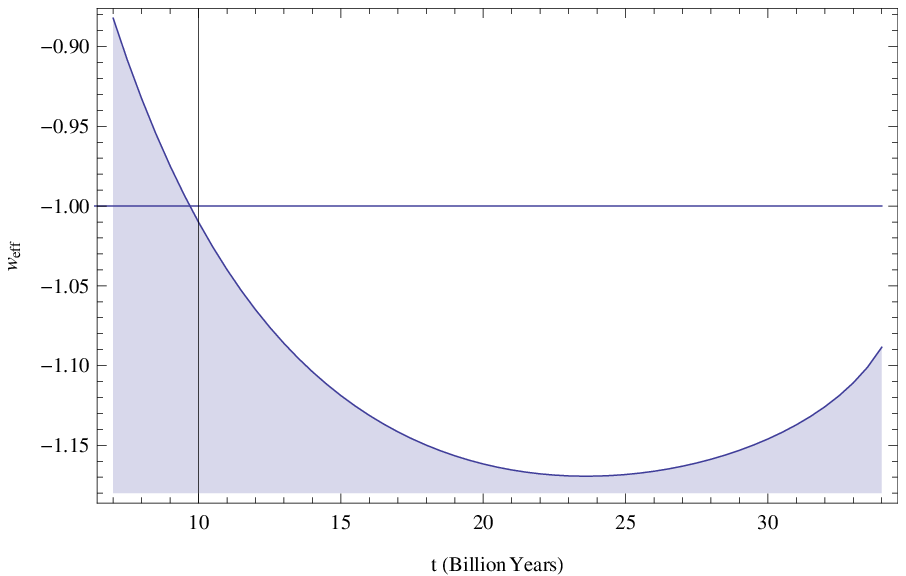}
\includegraphics[width=15pc]{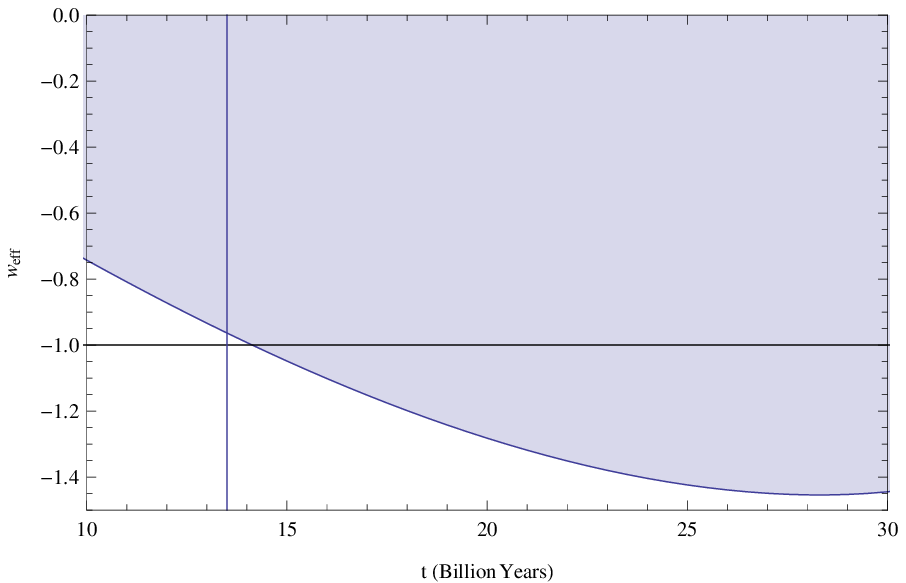}
\caption{The effective equation of state as a function of time (Gy). In the left plot, the parallel and the vertical line indicate the phantom divide line and the time instance that the crossing occurs. In the right plot the parallel and the vertical line indicate the phantom divide line and the present time $13.5$ billion years.}
\label{eosplot1}
\end{figure}

The deformed matter bounce model (\ref{deformedscalefactor}) has one more appealing feature and particularly it can realize a late-time acceleration era, as we now demonstrate. Particularly, if the parameters $\rho_c$ and $f_0$ are chosen to satisfy $\rho_c/f_0\sim 10^{-4}$ and also if $t_s\simeq 10^{18}$sec, then a numerical computation can show that the deceleration to acceleration transition occurs near $t\simeq 10^{17}$sec (4-5 billion years earlier from present time), and also after this transition, the Hubble rate is described by $H_s(t)$ solely, since it dominates over the matter bounce one $H_{mb}(t)$. In order to be more quantitative at this point, in Table \ref{tablei}, we present some characteristic values of the Hubble rates $H_{mb}(t)$ and $H_{s}(t)$, at late times, for $\alpha=4/3$, $t_s=10^{18}$sec and $\rho_c/f_0\simeq 10^{-4}$. As it can be seen, clearly the Hubble rate $H_{s}(t)$ dominates over $H_{mb}(t)$ for the values of the parameters being as specified.
\begin{table*}
\small
\caption{\label{tablei} The Values of the fraction $H_s(t)/H_{mb}(t)$, for various values of the cosmic time.}
\begin{tabular}{@{}crrrrrrrrrrr@{}}
\tableline
\tableline
\tableline
Hubble Rates & $t=10^{17}$sec & $t=5\times 10^{17}$sec & $t=9\times 10^{17}$sec
\\\tableline
$\frac{H_s(t)}{H_{mb}(t)}$ & 1726.24 & 18898.8 & 290849
\\\tableline
\tableline
 \end{tabular}
\end{table*}
In order to reveal the accelerating expansion at late times, let us compute the EoS $w_{eff}=P/\rho$,
\begin{equation}\label{eosdef}
w_{eff}=-1-\frac{\dot{H}}{3H^2}-\frac{\rho_c \dot{H}}{3 H^2 \sqrt{\rho_c \left(\rho_c-12 H^2\right)}}\, ,
\end{equation}
for the deformed Hubble rate of Eq. (\ref{deformedscalefactor}). The behavior of the EoS depends on the choice of the parameters for our model, but as we now show, the qualitative behavior is the same. Particularly, the EoS deforms from the matter domination value $w_{eff}=0$, to a quintessential value $-1<w_{eff}<0$, which describes the dark energy era preceding the matter domination era. Eventually, the Universe crosses the phantom divide line and is described by a phantom dark energy era, and this behavior persists until an era near the Rip. The crossing of the phantom divide line strongly depends on the choice of the parameters of our model, and in order to see this, in Fig. \ref{eosplot1}, we plotted the behavior of the EoS $w_{eff}$ as a function of the cosmic time in billion years. Note that the present era corresponds to approximately $13.5$ billion years, the time instance $t_s$ is assumed to occur at $35$ billion years ($\sim 10^{18}$sec), and also the parameters are appropriately adjusted.
In the left plot of Fig. (\ref{eosplot1}), the fraction $r_{l}=\rho_c/f_0$ is chosen to satisfy $r_l=0.3r_R$, where $r_{R}=\rho_c/f_0$ is the corresponding fraction for the values of the parameters corresponding to the right plot. As it can be seen, a slight change in the parameters changes the time instance that the crossing of the phantom divide line occurs. The qualitative picture is the same and only the time scales that the various events occur change. Particularly, in the left plot of Fig. (\ref{eosplot1}), the EoS crosses the phantom divide line approximately three billion years before the present time era, hence at present time the Universe is described by a phantom dark energy era. In the right plot, the crossing occurs at $14.5$ billion years, so the present time era is described by a quintessential dark energy era, which becomes phantom eventually after the present time era. 

In both cases, the very late time era is described by a phantom dark energy era. This can also be seen from the expression (\ref{eosdef}), owing to the fact that a Big Rip requires $\dot{H}>0$, and we discussed this issue previously, below Eq. (\ref{inertialforcecomov}). But we can explicitly show that indeed a phantom dark energy era occurs at very late times, since a direct calculation yields that, as $t\to t_s$, and during the dark energy era, the EoS becomes approximately equal to,
\begin{equation}\label{eos1}
w_{eff}\simeq -1+\frac{(t-t_s)^{-1+\alpha } \alpha }{3 f_0}+\frac{(t-t_s)^{-1+\alpha } \alpha  \rho_c}{3 f_0 \sqrt{\rho_c \left(\rho_c-12 \left(f_0 (t-t_s)^{-\alpha }+\frac{2 t \rho_c}{4+3 t^2 \rho_c}\right)^2\right)}}\, ,
\end{equation}
where we have substituted the Hubble rate of Eq. (\ref{deformedscalefactor}) in the expression for the EoS given in Eq. (\ref{eosdef}), and we have kept the dominant terms in the limit $t\to t_s$. Since $\alpha>1$ and also $\alpha=2n/(2m+1)$, the EoS is equal to, 
\begin{equation}\label{eos1fromlathos}
w_{eff}\simeq -1-\frac{(t_s-t)^{-1+\alpha } \alpha }{3 f_0}-\frac{(t_s-t)^{-1+\alpha } \alpha  \rho_c}{3 f_0 \sqrt{\rho_c \left(\rho_c-12 \left(f_0 (t-t_s)^{-\alpha }+\frac{2 t \rho_c}{4+3 t^2 \rho_c}\right)^2\right)}}\, ,
\end{equation}
so the cosmological evolution is described by a phantom evolution before the Rip. During the expansion phase of the matter bounce era, before the late-time singular deformed part of the Hubble rate controls the evolution, the EoS is approximately equal to $w_{eff}\simeq 0$, so the matter domination era precedes the late-time acceleration era. More importantly, before the Rip, and since  $\alpha=2n/(2m+1)$, the EoS (\ref{eos1}) describes a phantom acceleration era. This shows that in the present model, the matter dominated evolution precedes a quintessential dark energy era, which turns into a phantom dark energy evolution ending abruptly in a Rip singularity. This behavior agrees with the classical viewpoint, since it is known \cite{sergeirip} that usually a Rip singularity is realized only from a phantom dark energy era, and this behavior occurs in the LQC context too. Note however that the EoS also becomes complex near the Rip. A numerical computation for the values of the parameters we used earlier shows that nearly a second before the Rip, the energy density, the pressure and in effect the EoS become complex. Thus, the Rip cannot be reached by the present theoretical model. 

Hence with the singular deformed matter bounce scenario, we have the following features which are listed below:
\begin{itemize}
    \item At early times, at the beginning of the contracting phase, a scale invariant power spectrum is generated.
    \item The initial singularity is avoided.
    \item The infinite bouncing cycle stops when the Rip singularity at late times is reached-actually it is never reached and the physical description ceases to consistently describe the era near the Rip.
    \item The late-time acceleration era occurs, which is described by a quintessential dark energy era which turns into phantom, ending at a Rip singularity.
\end{itemize}   
Before proceeding, let us recapitulate here the qualitative features of the deformed matter bounce model. At the beginning of the cosmological evolution, early in the contracting phase, the Universe was approximately described by the matter bounce scale factor $a_{mb}(t)\sim t^{2/3}$, since the contribution from the deformation part was negligible. The Universe started contracting and bounced off at $t=0$, still being described by the matter bounce scenario. Then it continued its evolution, being described by the matter bounce scenario, and at late times, but well before the Rip singularity, the Universe evolved with scale factor $a_{mb}(t)\sim t^{2/3}$. Finally, the deformation part starts to control the evolution and the Universe starts to expand in an accelerating way. It can be shown that, if the parameters are appropriately chosen, the evolution can have quite appealing properties, since the deceleration-acceleration transition can occur at the phenomenologically acceptable time instance. Finally, at very late times, the Universe ends abruptly its evolution, since a Rip singularity occurs, and the infinite bouncing circle is avoided. With regards to the primordial perturbations issue, their behavior is approximately  identical to the standard matter bounce scenario perturbations, so a scale invariant spectrum is obtained, corresponding to the early-time era, well inside the contracting phase. Also we need to stress again that the LQC description of our model ceases to be valid at the vicinity of the Rip singularity, so this shows that possibly a more fundamental theory maybe controls the final stage of this phantom and catastrophic evolution.

\subsubsection{The Late-time Attractor}

As we already discussed, the deformed matter bounce scenario has as a late-time attractor which is a dark energy era. Moreover, the model we presented becomes unphysical at finite-time in the future, at which time instance, the scale factor and the Hubble rate diverge, but the energy density and effective pressure of the LQC model become complex. Therefore we claimed that this finite-time in the far future cannot be reached, at least in a physical way, since the complex values indicate that this lies beyond the approximations that hold true for the model. 

\begin{figure}[h] \centering
\includegraphics[width=15pc]{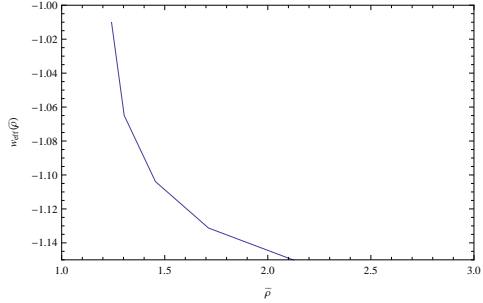}
\caption{The parametric plot of the effective equation of state $w_{eff}$ and of the rescaled energy density $\bar{\rho}$.}
\label{ploteos}
\end{figure}

However, we need to address this issue in a more quantitative way, since this qualitative picture has to be further supported for clarity. The purpose of this section is two-fold: firstly we will confirm that the late-time attractor of the LQC deformed matter bounce model is a dark energy era. Secondly, we need to support our claim that actually the physical system, quantified in terms of a dynamical system, cannot reach the Rip singularity which occurs in the far future. For our dynamical system analysis, we shall partially use the notation of Ref. \cite{nobigrip1}, and we introduce the following variables (note that we use a physical dimensions system for which $\kappa=1$):
\begin{equation}\label{variablesdyna}
x=\frac{\rho}{3H^2},\,\,\,y=\frac{\rho}{\rho_c}.
\end{equation}
Owing to Eq. (\ref{holcor1}), the variables $x$ and $y$ satisfy the relation,
\begin{equation}\label{xyrelation}
y=1-\frac{1}{x}\, ,
\end{equation}
and from (\ref{holcor1}), we can easily read off the constraints for $x$ and $y$, which are,
\begin{equation}\label{constraintxy}
x>1,\,\,\,0<y<1\, .
\end{equation}
At this point it is easy to see why it is not possible for the physical system at hand to reach a Rip singularity, since the variable $y$ would become complex at the time instance where the singularity occurs $t=t_s$, and also at the vicinity of the singularity, $y$ would take values $y>1$, before even it becomes singular. Hence, the path from the late-time attractor, which in our case as we demonstrate is a dark energy era, is blocked. Having made that clear, now we demonstrate that the late-time attractor is indeed a dark energy era. The variable $x$ satisfies the following dynamical equation,
\begin{equation}\label{lordequations}
\frac{\mathrm{d}x}{\mathrm{d}N}=-3\left[1+w_{eff}(y(x))\right]x^2\left ( 1-\frac{1}{x}\right)\, ,
\end{equation}
where we have made use of the $e$-foldings number $N=\ln a$, and also the function $w_{eff}(y (x))=P/\rho$, is the EoS, which is an explicit function of $y$ and an implicit function of $x$, via Eq. (\ref{xyrelation}). Also the variable $y$ satisfies the following differential equation,
\begin{equation}\label{diffeqny}
\frac{\mathrm{d}y}{\mathrm{d}\bar{t}}=-3 \left[ 1+w_{eff}(y)\bar{H} \right]y\, ,  
\end{equation}
and also,
\begin{equation}\label{yequation}
\bar{H}^2=y \left(1-y \right)\, ,
\end{equation}
with $\bar{H}=\frac{H}{H_c}$, $\bar{t}=H_c t$ and $H_c=\sqrt{\rho_c}{3}$. What will actually reveal the dynamical behavior of the model is the EoS function $w_{eff} (\rho)$, but unfortunately we cannot obtain this function, due to the lack of analyticity. What we have at hand are equations that yield the EoS as a function of the cosmic time, given in Eqs. (\ref{eosdef}) and (\ref{eos1fromlathos}), and also the energy density and the effective pressure as functions of the cosmic time. The only way to find an approximation of the function $w_{eff}(\rho)$ is to proceed numerically and find a convenient fit of the $w_{eff} (\rho)$ curve, by using the functions $w_{eff}(t)$ and $\rho (t)$ and by finding their parametric plot. Then a convenient fitting will yield the approximate form of the EoS $w_{eff}$. It is therefore vital to find the approximate behavior of the EoS as a function of the energy density (and implicitly in terms of the variable $y$). We shall focus on the case that the present era corresponds to the left plot of Fig. (\ref{eosplot1}), so the dark energy era is a phantom one, but the same applies if we choose the quintessential dark energy era. By making the parametric plot of the EoS $w_{eff}(t)$ and of the energy density $\bar{\rho}$, we obtain the curve appearing in Fig. \ref{ploteos}, where for convenience we introduced the variable $\bar{\rho}=\rho\times 10^{11}$, since the values of the energy density near the present time epoch are particularly small.
We can optimally fit the curve of Fig. \ref{ploteos}, by using the data of the values of $w_{eff}$ and the corresponding values of $\bar{\rho}$, and it is found that the EoS as a function of $\bar{\rho}$ behaves as follows,
\begin{equation}\label{eosasfunctionoft}
w_{eff}(\bar{\rho})=-1-c\,\bar{\rho}^{\beta}\, ,
\end{equation}
with $c=0.0754968$ and $\beta=0.548649$. Note that $\beta<1$, so this will play some role in the following analysis. Having the function $w_{eff}(\bar{\rho})$ at hand, we can easily find that the function $w_{eff}(x)$ is,
\begin{equation}\label{weffx}
w_{eff}(x)=-1-\mathcal{A}\rho_c^{\beta}\left( 1-\frac{1}{x}\right)^{\beta}\, ,
\end{equation}
where $\mathcal{A}=c\times 10^{11\beta}$. Since the variables $x$ and $y$ are interrelated via Eq. (\ref{xyrelation}), by investigating one of the two differential equations (\ref{lordequations}) and (\ref{diffeqny}), suffices to determine the behavior of the dynamical system (\ref{lordequations}), (\ref{diffeqny}) and (\ref{yequation}), so from now on we focus on the study of Eq. (\ref{lordequations}). Finding the asymptotic behavior of $x$ will effectively determine the asymptotic attractor $(x,y,\bar{H})$. By using Eq. (\ref{weffx}), the differential equation (\ref{lordequations}) becomes,
\begin{equation}\label{lordeeqfinal}
\frac{\mathrm{d}x}{\mathrm{d}N}=3\,\mathcal{A}\,\rho_c^{\beta}\,x^2\,\left(1-\frac{1}{x} \right)^{\beta+1}\, .
\end{equation} 
This equation, in conjunction with Eqs. (\ref{xyrelation}) and (\ref{yequation}), constitute the dynamical system which determines the dynamical evolution of the model we discuss. We can solve analytically the differential equation (\ref{lordeeqfinal}), the solution of which is,
\begin{equation}\label{soldiffeq}
\left( 1-\frac{1}{x} \right)^{-\beta}=-3\beta\, \mathcal{A}\rho_c^{\beta}\,N+\beta \, N_1\, ,
\end{equation}
where $N_1$ is an integration constant and recall that we previously found that $\beta<1$.

Here we shall demonstrate that the late-time attractor is indeed a dark energy era. An important remark is in order: The study we shall perform shortly does not require to fix the Hubble rate to be equal to the Hubble rate of the deformed matter bounce model of Eq. (\ref{deformedscalefactor}). This Hubble rate was used only in order to obtain the function $w_{eff} (x)$, namely Eq. (\ref{weffx}), and nothing else. From now on we shall assume that the variable $\bar{H}$ is a free variable of the system, and we shall demonstrate that the EoS of Eq. (\ref{weffx}) leads to a phantom dark energy late-time attractor, the characteristic values of which we shall confirm numerically, for the Hubble rate (\ref{deformedscalefactor}).

Having made that clear, we now proceed to find the fixed points of the dynamical system, and a fixed point of the differential equation (\ref{lordeeqfinal}) is $x=0$, which however is not a physical solution, since $y<0$ and it is singular in this case and also, this would mean that $\bar{H}$ also blows up since it depends on positive powers of $y$. Recall that this exactly describes the Rip case, which is an unphysical solution, so in some sense, the deformed matter bounce model, which corresponds to the EoS (\ref{weffx}), has a very late-time attractor, the Rip singularity, which however cannot be reached by any physical way. This unphysical behavior can be confirmed also by using the solution (\ref{soldiffeq}), which in the limit $x\to 0$ yields the unphysical result that $N$ becomes complex. Hence we do not further discuss about this unphysical fixed point. 

A physical fixed point of the dynamical system however is $x=1$, so by also taking into account Eqs. (\ref{xyrelation}) and (\ref{yequation}), one fixed point of the dynamical system is $(x,y,\bar{H})=(1,0,0)$, and we now discuss the implications of such a fixed point. From the solution (\ref{soldiffeq}), as $x\to 1$, this corresponds to large values of $N$, which can be achieved at very large values of the scale factor, so effectively this is the final attractor of the dynamical system. 

\begin{figure}[h] \centering
\includegraphics[width=15pc]{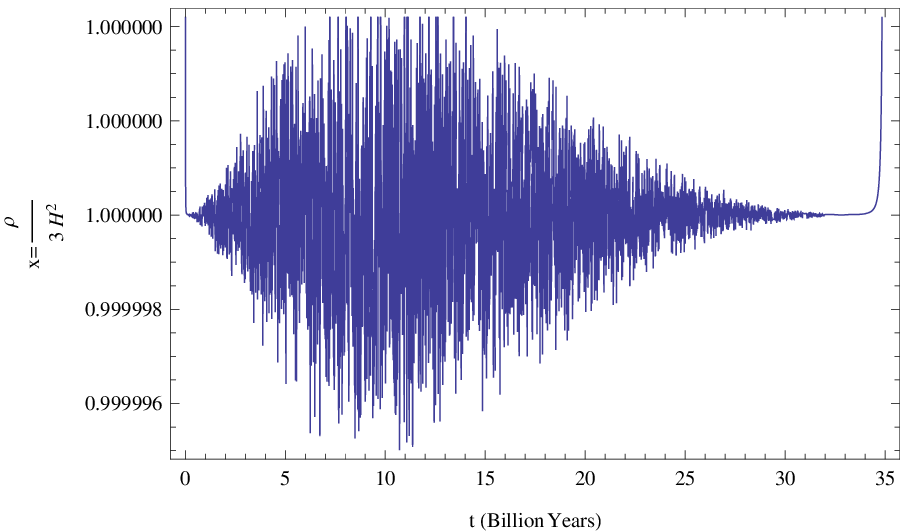}
\includegraphics[width=15pc]{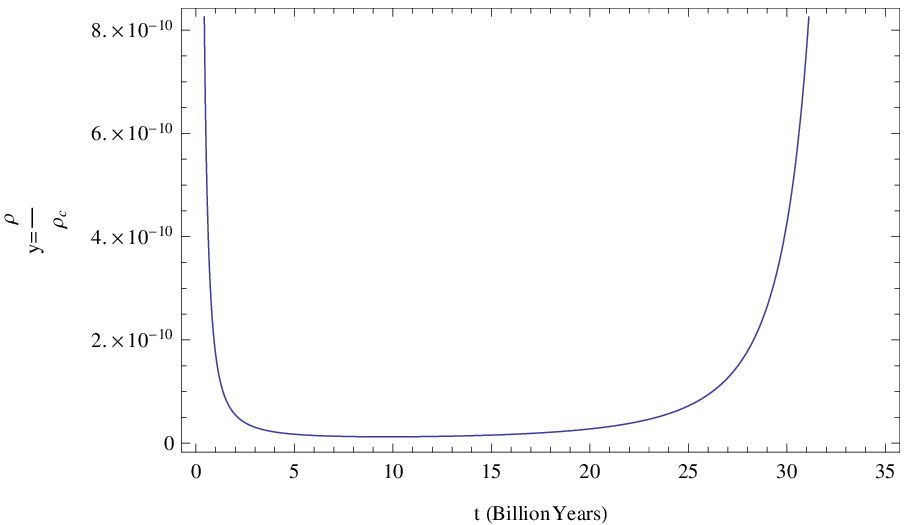}
\includegraphics[width=15pc]{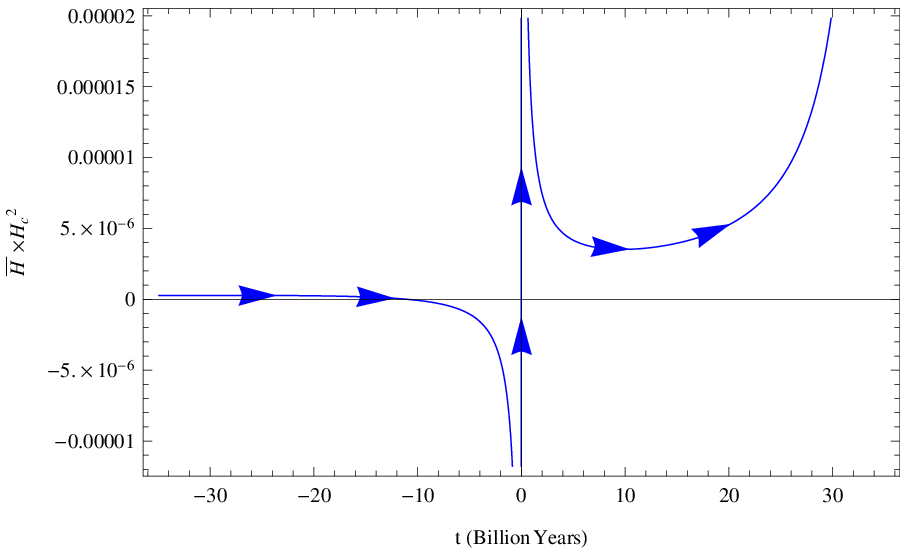}
\caption{The parameter $x$ (left plot), $y$ (right plot) and $\bar{H}$ (bottom plot), as functions of the cosmic time, for the deformed matter bounce model (\ref{deformedscalefactor}). It can be seen that the late-time attractor $(x,y,\bar{H})=(1,0,0)$ occurs near the present time $t\simeq 13.5 $ billion years.}
\label{eosfinalplots}
\end{figure}

Therefore the result of our analysis is that the late-time attractor of the dynamical system is the fixed point $(x,y,\bar{H})=(1,0,0)$, and by late-time, we mean cosmic times which do not render the physical system unphysical, with the latter case corresponding to the Rip time. Effectively, late-times for which the behavior $(x,y,\bar{H})=(1,0,0)$, qualify to be considered as late times, so now we shall numerically investigate if the deformed matter bounce model (\ref{deformedscalefactor}) at times near the present time era (13.5 billion years), generates such a behavior. In Fig. \ref{eosfinalplots}, we have plotted the parameter $x$ (left plot), $y$ (right plot) and $\bar{H}$ (bottom plot), as functions of the cosmic time, for the deformed matter bounce model (\ref{deformedscalefactor}).
As it can be seen from Fig. \ref{eosfinalplots}, the late-time attractor $(x,y,\bar{H})=(1,0,0)$ is reached near $t\sim 13.5$ billion years, so it actually occurs near the present time. Also, as we proved earlier and it can also be seen in Fig. (\ref{eosplot1}), the EoS near the present time corresponds to that of a phantom accelerating epoch, therefore the attractor $(x,y,\bar{H})=(1,0,0)$, actually describes a phantom dark energy attractor. However, as we already mentioned earlier, by appropriately choosing the parameters, the late-time attractor can be a quintessential dark energy era, but we deal with the first case here, since the qualitative picture is the same.

Before closing a final remark is in order. The solution (\ref{soldiffeq}) of the differential equation (\ref{lordeeqfinal}) in the limit $x\to \infty$, yields that $y\to 1$ and $\bar{H}\to 0$, and this is achieved at a finite $N$. Actually, the point $(x,y,\bar{H})=(\infty ,1,0)$ is the bouncing point, at which $\bar{H}\to 0$ and $y$ tends to a finite value, so the effective energy density is actually finite at this point.

\subsection{The Question of Viability and the Cosmological Evolution in the Deformed Matter Bounce Model}

An interesting question which we need to address before we close this section is whether this model is viable or not. This is a deep question and the quick answer is that the deformed matter bounce model is as viable as the ordinary matter bounce model is, with the difference between these two models being that the deformed matter bounce model generates a late-time acceleration era. In this section we shall address in detail this issue, but before getting started, let us point out that the matter bounce model is not an inflationary model, so it has no period of inflation. This is a very important observation and this will affect the interpretation of the Universe evolution. So the question if the model is viable or not has a simple answer, it is as viable as the matter bounce model, and therefore the power spectrum of primordial curvature perturbations is scale invariant and also the early Universe, which corresponds to the contraction epoch, is in an non-causal state. This is the drawback of the matter bounce model and the deformed model also shares this rather not so appealing feature.

\begin{figure}[h] \centering
\includegraphics[width=15pc]{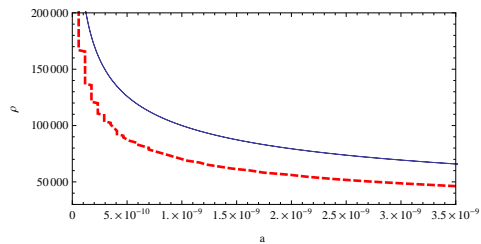}
\caption{Comparison of the parametric plots (left up plot) of the function $\rho (t)$ versus the function $a(t)$, for the deformed matter bounce (red, dashed curve) and the ordinary matter bounce scenario (blue curve). We have used the values $\alpha=4/3$, $t_s=10^{18}$sec and $\rho_c/f_0\simeq 10^{-4}$ and the parametric plot corresponds to the time interval $10^{17}<t<10^{18}$ sec.}
\label{revisionplotrhoalpha}
\end{figure}

The dynamics of the deformed model can be represented perfectly by plotting the $\rho-a$ graph, however in order to find the function $ \rho ( a)$, we need to invert the function $a (t)$, which for the scale factor (\ref{deformedscalefactor}) is not possible to do analytically. However, we can make the parametric plot for the functions $\rho (t)- a(t)$ which appears in Fig. \ref{revisionplotrhoalpha}, which corresponds to the whole time interval $10^{17}<t<10^{18}$ sec, and for the values of the parameters chosen as $\alpha=4/3$, $t_s=10^{18}$sec and $\rho_c/f_0\simeq 10^{-4}$.
As it can be seen in Fig. \ref{revisionplotrhoalpha}, the two models differ clearly in these plots, after approximately $t\simeq 10^{17}$sec, where the deformed matter bounce deviates from the behavior $\rho\sim a^{-3}$.

However in the absence of an analytic expression for the function $\rho(a)$, we shall provide another viewpoint of the comparison of the two models. Particularly we shall compare the Hubble radii $R_H(t)=1/(a(t)H(t))$ of the two models as functions of the cosmic time. But why choosing the Hubble radius as a measure of comparison? The answer is that the Hubble radius can be used in order to determine the physical viability and appeal of a cosmological model. The Hubble radius in an inflationary cosmology behaves in a different way in comparison to a bounce model. This is because in the bounce cosmology case, a contracting phase occurs too. In the case of the matter bounce, the Hubble horizon during the contraction era shrinks, and eventually the sub-horizon primordial modes exit the Hubble radius. These modes are actually the ones responsible for the generation of the scale invariant power spectrum \cite{brande1}.   
\begin{figure}[h] \centering
\includegraphics[width=15pc]{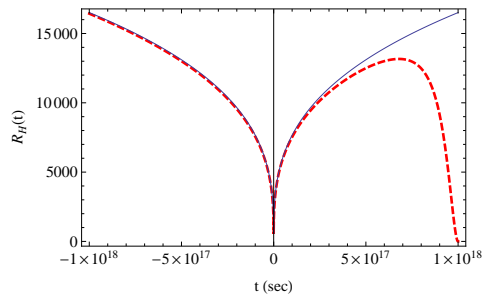}
\caption{Comparison of the Hubble radius $R_H(t)=1/(a(t)H(t))$ as function of the cosmic time $t$, for the deformed matter bounce (red, dashed curve) for the ordinary matter bounce scenario (blue curve). We have used the values $\alpha=4/3$, $t_s=10^{18}$sec and $\rho_c/f_0\simeq 10^{-4}$.}
\label{hubbleraddii}
\end{figure}
Eventually, after the bounce, the Hubble horizon expands and this actually corresponds to the matter dominated epoch. Hence the primordial model re-enter the horizon at some time instance during the expansion of the Hubble radius. In the standard matter bounce scenario, the expansion of the Hubble horizon never stops, so the horizon increases its size. Hence the late-time acceleration era cannot be described, since during the late-time acceleration era, the Hubble radius would shrink again. One appealing feature of the deformed matter bounce model is that after the end of the matter domination era, the Hubble radius shrinks again, and thus the late-time occurs. In Fig. \ref{hubbleraddii} we compared the behavior of the Hubble radius $R_H(t)$ as a function of the cosmic time $t$, for both the deformed and ordinary matter bounce models. The red and dashed curve corresponds to the deformed matter bounce model and the blue to the ordinary. As it can be seen, in the deformed model, the Hubble radius shrinks around $t\sim \times 10^{17}$sec, and therefore the late-time era can be consistently described. Thus the deformed model has all the features of the matter bounce model (regardless if these are appealing or not), for all the eras except at late times, where it differs significantly.

\section{The Viscous Matter Fluid and Canonical Scalar Field Description Viewpoint}

In the previous we demonstrated that we can have a nearly matter bounce scenario with a late-time dark energy era and a Rip singularity at the end of time, and in order to achieve this, we needed to deform the matter bounce scenario, so that the deformation part actually alters the late-time behavior, and therefore affecting only the early-time behavior. As we showed, this behavior can occur if the total matter energy density consists of two parts, and is of the form $\rho=\rho_{mb}+\rho_s$, with $\rho_{mb}$ describing a pressureless matter fluid and $\rho_s$ describing a matter fluid with pressure $P_s$. In this section we shall try to formalize this assumption for the effective energy density further, and we shall use two different theoretical frameworks which can support such a matter energy density behavior, namely, the effective viscous fluid description \cite{newref1,newref2,sergeiperf1,sergeiperf2,sergeiperf3,sergeiperf4} and also the canonical scalar field description \cite{polonos}.

\subsection{Viscous Matter Fluid Approach}

With regards to the imperfect fluid approach, it is quite usual in the context of modified gravity to use viscous fluids in order to realize various cosmological scenarios, see for example \cite{newref1,newref2,sergeiperf1,sergeiperf2,sergeiperf3,sergeiperf4} for some interesting cosmological realizations. In the present paper we shall be interested in describing the deformed matter bounce scenario, in terms of a viscous fluid with bulk viscosity $B(a(t),H,\dot{H},..)$. As it can be seen, in general, the bulk viscosity function can be a function of the scale factor, the Hubble rate and the higher derivatives of the latter. We investigate what is the EoS of the viscous fluid that describes the deformed matter bounce scenario. We shall make use of the FRW equations of motion for LQC, namely Eqs. (\ref{holcor1}) and (\ref{eqnm}), which are satisfied by the total energy density $\rho$ and by the pressure $P$. Recall that the pressure is due to the extra deformation part $\rho_s$ in the energy density. By solving Eq. (\ref{eqnm}) with respect to $P$, we get,
\begin{equation}\label{metviw}
P=-\rho-\frac{2\dot{H}}{1-\frac{2\rho}{\rho_c}}\, ,
\end{equation}
and owing to Eq. (\ref{holcor1}), the effective pressure of the viscous fluid becomes,
\begin{equation}\label{effpress}
P=-\rho-\frac{2\dot{H}}{1-\frac{\rho_c-\sqrt{-12 H^2 \rho_c+\rho_c^2}}{\rho_c}}\, .
\end{equation}
The equation (\ref{effpress}) is actually the EoS of the viscous fluid that can realize the deformed matter bounce cosmology of (\ref{deformedscalefactor}) and in view of Eq. (\ref{limits}), it describes for almost all the cosmic eras, a matter dominated Universe, which reveals its singulary deformed character only at late times. From the EoS (\ref{effpress}), the bulk viscosity function $B(H,\dot{H})$ can easily be read off, and it is equal to,
\begin{equation}\label{bulviscosity}
B(H,\dot{H})=\frac{2\rho_c\dot{H}}{\sqrt{-12 H^2 \rho_c+\rho_c^2}}\, .
\end{equation}
Notice that for the cosmological evolution (\ref{deformedscalefactor}), the bulk viscosity (\ref{bulviscosity}) is positive (recall that $\dot{H}>0$ from the previous section), and therefore it satisfies the important requirement that constraints the entropy change to have a positive sign during irreversible processes \cite{sergeiperf1,sergeiperf2,sergeiperf3,sergeiperf4}. Finally, in view of the imperfect fluid EoS, the energy momentum tensor of the viscous fluid is,
\begin{equation}\label{viscousenergymom}
T_{\mu \nu}=\rho\,u_{\mu}\,u_{\nu}+\Big{(}\rho+B(H,\dot{H})\Big{)}\Big{[}g_{\mu \nu}+u_{\mu}\,u_{\nu}\Big{]}\, ,
\end{equation}
where $B(H,\dot{H})$ is the bulk viscosity and $u_{\mu}$ is the comoving four velocity, which for the FRW background metric (\ref{metricfrw}) reads $u_{\mu}=(1,0,0,0)$.

\subsection{Canonical Scalar Field Description}

From another viewpoint, we can use a canonical scalar field as being the source of the energy density $\rho$, so now we investigate the LQC description of a canonical scalar field $\phi$, the pressure ($P$) and the energy density ($\rho$) of which are,
\begin{equation}\label{rhopre}
\rho=\frac{\dot{\phi}^2}{2}+V(\phi ),\,\,\,P=\frac{\dot{\phi}^2}{2}-V(\phi )\, ,
\end{equation}
with $V(\phi)$ being the scalar potential of the canonical scalar field. In Refs. \cite{polonos,matterbounce7}, a similar presentation was given for the specific equation of state $P=w\rho$ obeyed by the pressure and the energy density of the scalar field, but here we shall work with the equation of state given in Eq. (\ref{effpress}). By using Eq. (\ref{rhopre}), the potential and the scalar fields as functions of time are equal to,
\begin{equation}\label{potphi}
V(\phi (t))=\rho(t)-P(t)=2 \rho (t)+\frac{2\dot{H}}{1-\frac{2\rho (t)}{\rho_c}}\, ,
\end{equation}
\begin{equation}\label{phiovertin}
\phi (t)=\pm \int \sqrt{-\frac{2\dot{H}}{1-\frac{2\rho (t)}{\rho_c}}}\,\mathrm{d}t\, ,
\end{equation}
hence, given the Hubble rate, by solving (\ref{phiovertin}) with respect to the cosmic time $t$ and substituting in Eq. (\ref{potphi}), the potential $V(\phi)$ easily follows. For the Hubble rate of Eq. (\ref{deformedscalefactor}), the potential reads,
\begin{align}\label{pot1}
& V(\phi (t))=\rho_c+\frac{2 (t-t_s)^{-1-\alpha } \rho_c \left(2 (t-t_s)^{1+\alpha } \rho_c \left(-4+3 t^2 \rho_c\right)+f_0 \alpha  \left(4+3 t^2 \rho_c\right)^2\right)}{\left(4+3 t^2 \rho_c\right)^2 \sqrt{\rho_c \left(\rho_c-12 \left(f_0 (t-t_s)^{-\alpha }+\frac{2 t \rho_c}{4+3 t^2 \rho_c}\right)^2\right)}}\\ \notag &
+\sqrt{\rho_c \left(\rho_c-12 \left(f_0 (t-t_s)^{-\alpha }+\frac{2 t \rho_c}{4+3 t^2 \rho_c}\right)^2\right)}\, ,
\end{align}
however it is not easy to perform the integration of Eq. (\ref{phiovertin}) in an analytic way, so we examine the problem in the two limiting cases, that is, for times that the matter bounce scenario dominates in the Hubble rate (\ref{deformedscalefactor}) and for late times. Also in the next section we address this issue by using numerical analysis. We start off with the first scenario, which corresponds to times $t\ll t_s$, so the energy density is equal to,
\begin{equation}\label{rhosolasxeto}
\rho (t) \simeq \frac{\rho_c}{\frac{3}{4}t^2+1}\, ,
\end{equation}
and therefore the pressure $P$ appearing in Eq. (\ref{effpress}) is equal to zero, as was expected, since in this limit the physical system is described by the matter bounce. Accordingly, the potential $V(t)$ is equal to,
\begin{equation}\label{potnew1aaa}
V(t)\simeq \frac{4 \rho_c}{4+3 t^2 \rho_c}\, ,
\end{equation}
while the function $\phi (t)$ reads,
\begin{equation}\label{phit}
\phi (t) \simeq \frac{2 \sqrt{\frac{\rho_c}{4+3 t^2 \rho_c}} \sqrt{4+3 t^2 \rho_c}\, \,\mathrm{arcsinh} \left[\frac{1}{2} \sqrt{3} t \sqrt{\rho_c}\right]}{\sqrt{3} \sqrt{\rho_c}}\, ,
\end{equation}
where we have set the integration constant equal to zero, and by solving (\ref{phit}) with respect to $t$, and substituting in Eq. (\ref{potnew1aaa}), the resulting form of the potential is,
\begin{equation}\label{Vphi}
V(\phi) \simeq \rho_c\, \mathrm{sech}^2\left(\frac{\sqrt{3} \phi }{2}\right)\, ,
\end{equation}
which is identical to the potential found in Ref. \cite{polonos} (see also \cite{matterbounce7}) where the case with general equation of state parameter $w$ is studied\footnote{By making the comparison, we need to note that $\kappa^2=8\pi G$ in Ref. \cite{polonos}, and of course that $w=0$.}. At late times the effective energy density can be approximated by,
\begin{equation}\label{rhosolasxeto}
\rho (t)\simeq \frac{1}{2} \left(\rho_c-\sqrt{\rho_c \left(-12 f_0^2 (t-t_s)^{-2 \alpha }+\rho_c\right)}\right)\, ,
\end{equation}
since the deformation part in the scale factor dominates and therefore the corresponding pressure reads,
\begin{equation}\label{pressurenewdance}
P(t)\simeq \frac{1}{2} \left(-\rho_c+\sqrt{\rho_c \left(-12 f_0^2 (t-t_s)^{-2 \alpha }+\rho_c\right)} \left(-1-\frac{4 f_0 (t-t_s)^{-1+\alpha } \alpha }{-12 f_0^2+(t-t_s)^{2 \alpha } \rho_c}\right)\right)\, ,
\end{equation}
and therefore the corresponding potential $V(t)$ reads,
\begin{equation}\label{potnew1aaadanc}
V(t)=\rho_c+\sqrt{\rho_c \left(-12 f_0^2 (t-t_s)^{-2 \alpha }+\rho_c\right)} \left(1+\frac{2 f_0 (t-t_s)^{-1+\alpha } \alpha }{-12 f_0^2+(t-t_s)^{2 \alpha } \rho_c}\right)\, .
\end{equation}
The integral of Eq. (\ref{phiovertin}), cannot be done analytically, however by approximating the integrant appropriately, we get,
\begin{equation}\label{phitdance}
t-t_s\simeq \mathcal{A}\phi ^{-\frac{2}{-1+\alpha }},\,\,\,\mathcal{A}=2^{\frac{3}{-1+\alpha }} \left(\frac{(\alpha -1)}{\sqrt{f_0 \alpha  }}\right)^{-\frac{2}{-1+\alpha }}\, .
\end{equation}
Finally, by substituting in Eq. (\ref{potnew1aaa}), the potential reads,
\begin{equation}\label{Vphidance}
V(\phi)\simeq \rho_c+\sqrt{\rho_c \left(-12 f_0^2 \mathcal{A}^{-2 \alpha }\phi ^{-\frac{4\alpha }{-1+\alpha }}+\rho_c\right)} \left(1+\frac{2 f_0 \mathcal{A}^{-1+\alpha }\alpha  \phi ^{-\frac{2 (\alpha -1)}{-1+\alpha }}}{-12 f_0^2+\mathcal{A}^{-2 \alpha }\rho_c \phi ^{\frac{4\alpha }{-1+\alpha }}}\right)\, ,
\end{equation}
which can be further approximated as follows, 
\begin{equation}\label{powerlawphi}
V(\phi)\simeq 2 \rho_c-\frac{\rho_c \alpha }{6 f_0}\phi ^{-\frac{2 (\alpha -1)}{-1+\alpha }}\, ,
\end{equation}
hence at late times it is described by a nearly power law behavior in terms of the canonical scalar field $\phi$.

\subsubsection{Numerical Analysis for the Scalar Field}

\begin{figure}\centering
\includegraphics[width=15pc]{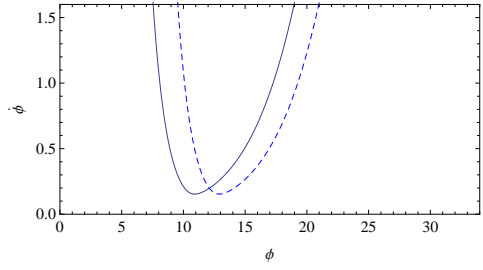}
\caption{The phase space diagrams $\dot{\phi}-\phi$, for the initial condition $\phi (0)=0.01$ (dashed), and for $\phi (0)=10$.}
\label{koukli}
\end{figure}

\begin{figure}\centering
\includegraphics[width=15pc]{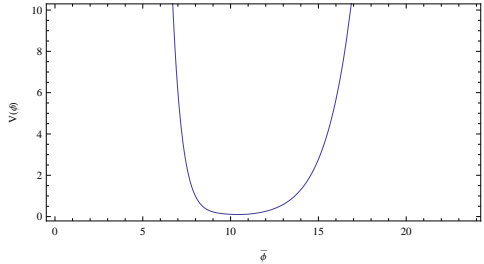}
\caption{The potential $V(\phi)$ as a function of $\phi$.}
\label{vff}
\end{figure}

As we discussed in the previous section, the LQC scalar potential of Eq. (\ref{pot1}), which corresponds to the deformed matter bounce scenario (\ref{deformedscalefactor}), cannot be expressed in terms of the canonical scalar field $\phi$, at least analytically. Hence, it is not possible to find the EoS as a function of $\phi$, at least analytically, therefore in this section we shall perform a numerical analysis in order to find an optimal approximation of the potential $V(\phi)$. Also we shall perform a phase space analysis of the scalar field $\phi (t)$, and we shall investigate its behavior as a function of the cosmic time. 
\begin{figure}[h] \centering
\includegraphics[width=12pc]{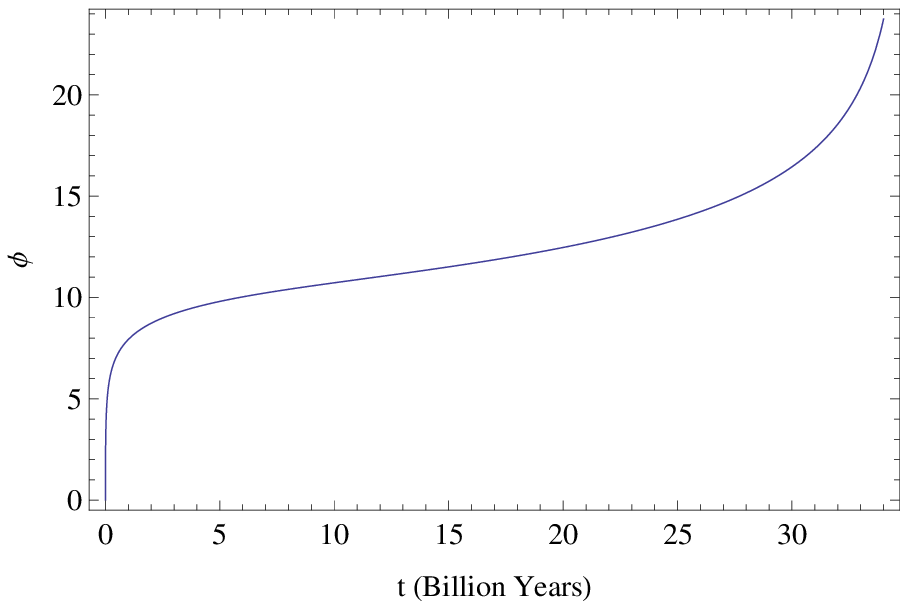}
\includegraphics[width=12pc]{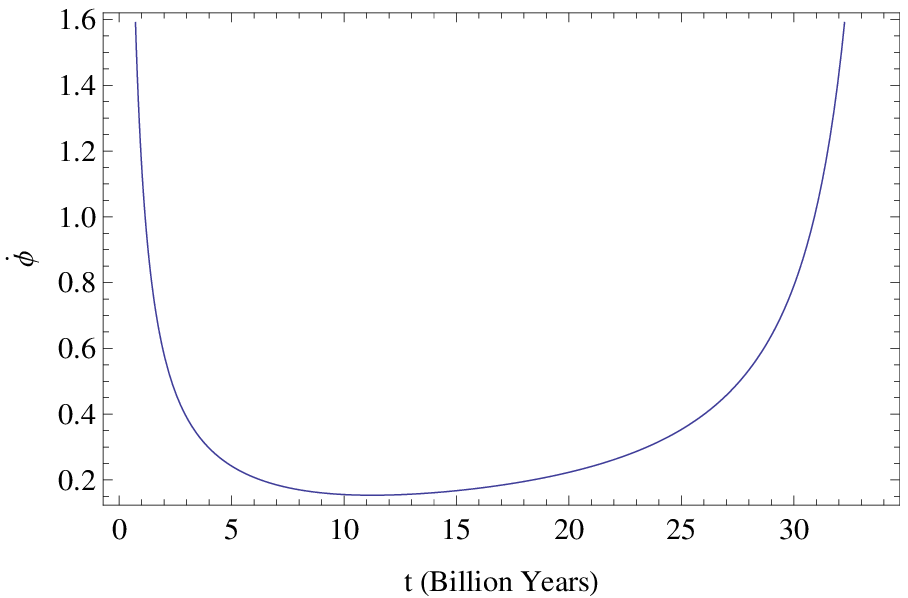}
\includegraphics[width=12pc]{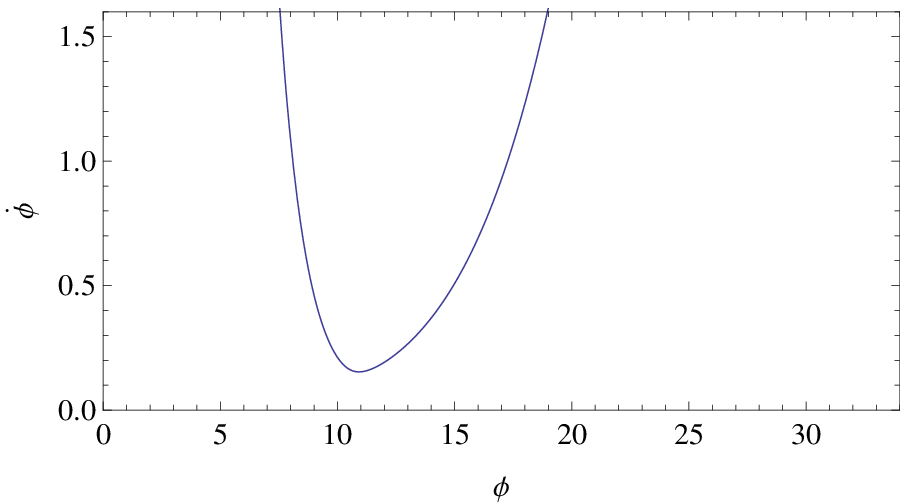}
\caption{The time dependence of the functions $\phi (t)$ (left), $\dot{\phi}(t)$ (right) and also the phase space diagram $\dot{\phi}-\phi$ (bottom), for the initial condition $\phi (0)=0.01$.}
\label{evolscal}
\end{figure}
This study will reveal how the canonical scalar theory evolves in time. We start off with the phase space analysis, and we solve numerically the differential equation (\ref{phiovertin}), by using various initial conditions. In Fig. \ref{evolscal}, we plotted the evolution of the functions $\phi (t)$ (left) and $\dot{\phi} (t)$ (right) as a functions of the cosmic time,  and also we included the phase space parametric plot $\dot{\phi}-\phi$ (bottom). As it can be seen, the scalar field during the present time era, that is $13.5$ billion years, varies very slowly in time. Note that the qualitative picture does not change if we alter the initial conditions, as it can be seen in Fig. \ref{koukli}.
Having a numerical solution for $\phi (t)$ and also knowing the time dependence of the scalar potential $V(t)$, enables us to find the parametric plot of the functions $V(\phi (t))-\phi (t)$, which appears in Fig. \ref{vff}. 
By optimally fitting the resulting curve, by using the data of the functions $V(\phi (t))$ and $\phi (t)$, we find that the curve of Fig. \ref{vff}, can be approximated by the following function $V(\phi )$,
\begin{equation}\label{vphi}
V(\phi)=c_1+c_2 \phi+c_3 \phi^2\, ,
\end{equation}
with $c_1\simeq 705$, $c_2\simeq 106$, $c_3\simeq 3.8$, and note that we focused for cosmic times in the interval $(13,15)$ billion years. Note that for the figures (\ref{vff}) and (\ref{koukli}) we have used the values $\alpha\to 4/3$, $\rho_c/f_0\sim 10^{9}$ and $t_s=35$ Billion years. The values for the parameters $c_i$, $i=1,2,3$ come from the fitting of the curve appearing in Fig. (\ref{vff}).

Having an approximation for the potential $V(\phi)$, and also the function $\phi (t)$, we can evaluate the EoS for the scalar field $w^{\phi}_{eff}$, which is defined to be,
\begin{equation}\label{nnnnnnnnnn}
w^{\phi}_{eff}=\frac{\frac{\dot{\phi}(t)^2}{2}-V(\phi (t))}{\frac{\dot{\phi}(t)^2}{2}+V(\phi (t))}\, ,
\end{equation}
and we can directly compare to the values of the EoS corresponding to the functional form of the EoS given in Eq. \ref{eos1}, which is the EoS (\ref{eosdef}) evaluated for the deformed matter bounce scenario. In Table \ref{tab1}, we have gathered the values for the EoS $w_{eff}$ and the scalar field EoS $w^{\phi}_{eff}$, for various cosmic times, and as it can be seen, the differences occur at the second decimal point of the EoS, which is a rather good estimation. 
\begin{table*}[h]
\small
\caption{\label{tab1}The values of the equation of state functions $w^{\phi}_{eff}$ and $w_{eff}$, for various cosmic times}
\begin{tabular}{@{}crrrrrrrrrrr@{}}
\tableline
\tableline
\tableline
Cosmic time (Billion Years) & $13$Gy & $13.1$Gy & $13.2$Gy & $13.5$Gy
\\\tableline
 $w^{\phi}_{eff}$ & -1.02998 & -1.01998 &  -1.01721 & -1.01065
\\\tableline
$w_{eff}$  &  -1.0861  & -1.08802 & -1.0899 &  -1.09538 
\\
\tableline
\tableline
 \end{tabular}
\end{table*}
Also notice that for the time interval $(13,13.5)$, the values of the EoS describe a phantom evolution, which agrees with the qualitative picture described in section II. In conclusion, the canonical scalar field description of the deformed matter bounce model (\ref{deformedscalefactor}) results to a phantom late-time attractor, with an EoS which is very close to the de Sitter value $-1$. Note however that, as in the description of the previous section, if the parameters are appropriately chosen, the EoS of the present time could be of quintessential type, which could turn to phantom after the present time era.

\section{$F(R)$ Gravity Description of the Singular Matter Bounce Scenario}

In this section we shall investigate which $F(R)$ gravity \cite{reviews1,reviews1a,reviews2,reviews3,reviews4} can describe the cosmological evolution (\ref{deformedscalefactor}). Obviously finding a complete analytic solution is a formidable task, since the scale factor and the corresponding Hubble rate are complicated, but we can obtain analytic results in various limits of the cosmic time. Actually, in a previous work \cite{matterbounce4} we found the vacuum $F(R)$ gravity which can realize the matter bounce scenario, for cosmic times near and away from the bounce, with the latter case corresponding to a scale factor analogous to $t^{2/3}$. As we demonstrated \cite{matterbounce4}, the vacuum $F(R)$ gravity that realizes the scale factor $a(t)\sim t^{2/3}$ is $F(R)\sim c_1 R^{\rho_1}+c_2R^{\rho_2}$, where $c_i$, $i=1,2$ are constants and $\rho_{1,2}$ are equal to $\rho_1=27/2$ and $\rho_2=-1/2$. Also near the bouncing point the $F(R)$ gravity is approximately equal to $F(R)\sim R^{\alpha}$, with $\alpha=\frac{-51-\sqrt{2433}}{2}$. Since the deformed matter bounce model of Eq. (\ref{deformedscalefactor}) is identical to the matter bounce for almost all eras, except near the Rip singularity, the $F(R)$ gravity in the corresponding limits behaves in the same way. What remains is to find the behavior of the $F(R)$ gravity for cosmic times near the Rip singularity. As we will demonstrate, for $\alpha=4/3$, the solution has a particularly interesting and well known functional form, which is that of an $R+\xi R^2$ gravity, with $\xi$ having a large value, in contrast to the standard $R^2$ inflation model \cite{starobinsky}, but we need to stress that this is just the leading order result and not the full $F(R)$ gravity description. As it is known, many cosmologies compatible with the $\Lambda$CDM model and with local astrophysical data, may be approximated in this way at leading order, like for example the Hu-Sawicki model \cite{sawicki}.

 In the context of $F(R)$ gravity it is possible to realize various cosmologies which were impossible to realize in Einstein-Hilbert gravity, unless certain assumptions were made. For instance bouncing cosmologies can occur in Einstein-Hilbert gravity only if the energy conditions are modified \cite{brande1}, but in the case of modified gravity the energy conditions are not affected. Here we will be interested for a vacuum theory, and by using well known reconstruction schemes \cite{Nojiri:2006gh}, we will investigate which $F(R)$ gravity can approximately realize the late-time evolution of the cosmological model (\ref{deformedscalefactor}), with emphasis given for times near the late-time Rip singularity. The vacuum four dimensional $F(R)$ gravity action is,
\begin{equation}
\label{action1dse}
\mathcal{S}=\frac{1}{2\kappa^2}\int \mathrm{d}^4x\sqrt{-g}F(R)\, .
\end{equation}
We adopt the metric formalism, in the context of which we vary the Jordan frame action of Eq. (\ref{action1dse}), with respect to the metric $g_{\mu\nu}$, and in effect the Friedmann equations read,
\begin{equation}
\label{frwf1}
 -18\left ( 4H(t)^2\dot{H}(t)+H(t)\ddot{H}(t)\right )F''(R)+3\left
[H^2(t)+\dot{H}(t)
\right ]F'(R)-
\frac{F(R)}{2}=0\, .
\end{equation}
We can rewrite the action of Eq. (\ref{action1dse}), in terms of an auxiliary non-dynamical scalar field $\phi$ as follows,
\begin{equation}
\label{neweqn123}
S=\int \mathrm{d}^4x\sqrt{-g}\left [ P(\phi )R+Q(\phi ) \right ]\, .
\end{equation}
A crucial point in our analysis is to find the functions $P(\phi (R) )$ and $Q(\phi (R) )$, so we vary the action (\ref{neweqn123}) with respect to $\phi$, and we acquire,
\begin{equation}
\label{auxiliaryeqns}
P'(\phi )R+Q'(\phi )=0\, ,
\end{equation}
where the prime indicates differentiation with respect to the auxiliary scalar $\phi$. As was shown in \cite{Nojiri:2006gh}, the scalar field $\phi$ and the cosmic time can be identified, so from now on we shall assume $t=\phi$. Solving the algebraic equation (\ref{auxiliaryeqns}), we obtain the function $t (R)$, so by substituting the result into Eq. (\ref{neweqn123}), we obtain the final form of the $F(R)$ gravity, which is,
\begin{equation}
\label{r1}
F(\phi( R))= P (\phi (R))R+Q (\phi (R))\, .
\end{equation}
Hence, finding the functions $P(\phi )$ and $Q(\phi )$ is a vital feature of the reconstruction method, and we can find the differential equations that these obey, by simply writing the equations of motion (\ref{frwf1}) in terms of these, and we obtain,
\begin{align}
\label{r2}
 & -6H^2P(\phi (t))-Q(\phi (t) )-6H\frac{\mathrm{d}P\left (\phi (t)\right
)}{\mathrm{d}t}=0\, , \nn
& \left ( 4\dot{H}+6H^2 \right ) P(\phi (t))+Q(\phi (t)
)+2\frac{\mathrm{d}^2P(\phi
(t))}{\mathrm
{d}t^2}+\frac{\mathrm{d}P(\phi (t))}{\mathrm{d}t}=0\, .
\end{align}
\begin{equation}
\label{r3}
2\frac{\mathrm{d}^2P(\phi (t))}{\mathrm {d}t^2}-2H(t)\frac{\mathrm{d}P(\phi
(t))}{\mathrm{d}t}+4\dot{H}P(\phi (t))=0\, .
\end{equation}
Having in mind the differential equation (\ref{r3}), given the Hubble rate, we can find in a straightforward way which $F(R)$ gravity actually realizes the given Hubble rate. Let us employ the method in the case that the Hubble rate is approximately equal to,
\begin{equation}\label{approxhub}
H(t)\simeq f_0( t-t_s)^{\alpha}\, ,
\end{equation}
so the differential equation of Eq. (\ref{r3}) becomes in this case,
\begin{equation}
\label{ptdiffeqn}
2x^{1+\alpha}\frac{\mathrm{d}^2P(x)}{\mathrm {d}x^2}-2f_0 x\frac{\mathrm{d}P(
t)}{\mathrm{d}t}-4f_0\alpha P(t)=0\, ,
\end{equation}
where we introduced the parameter $x=t-t_s$ and it is conceivable that as the Rip singularity is approached, this parameter tends to zero. The case of a general $\alpha$ is quite difficult to treat analytically, so we need to specify the parameter $\alpha$ in order to proceed. We shall present to illustrative examples, which we need to note that will yield the $F(R)$ gravity at leading order, so the result is an approximation and not the full solution and as we show, these forms of leading order of $F(R)$ gravities may originate from functions that are compatible with local astrophysical constraints.

We start off with the case $\alpha=4/3$, in which case the analytic solution of the differential equation (\ref{ptdiffeqn}) is the following,
\begin{align}
\label{genrealsola1}
& P(x)= \Big{(}1+6 f_0 \left(\frac{1}{x}\right)^{1/3}+\frac{63}{5} f_0^2 \left(\frac{1}{x}\right)^{2/3}+\frac{27}{4} f_0^4 \left(\frac{1}{x}\right)^{4/3}\\ \notag &
+\frac{81}{40} f_0^5 \left(\frac{1}{x}\right)^{5/3}+\frac{81 f_0^7 \left(\frac{1}{x}\right)^{7/3}}{2800}+\frac{243 f_0^8 \left(\frac{1}{x}\right)^{8/3}}{246400}+\frac{27 f_0^6}{80 x^2}+\frac{63 f_0^3}{5 x}\Big{)}C_1+C_2G_{22}\Big{(}z|9,-3{\,}^0\Big{)}
\, ,
\end{align}
with $C_1$, $C_2$ are arbitrary integration constants and the function $G_{pq}\Big{(}z|a,b{\,}^c\Big{)}$ is the Meijer function. In order to obtain an analytic solution, we can approximate $P(x)$ for $x\to 0$,
\begin{equation}\label{genrealsol}
P(x)\simeq \frac{243 C_1 f_0^8}{246400 x^{8/3}}+C_1\, ,
\end{equation}
and accordingly the function $Q(x)$ reads,
\begin{align}
\label{qtanalyticform}
Q(x) =-\frac{729 C_1 f_0^{10}}{123200 x^{16/3}}\, .
\end{align}
By substituting Eqs. (\ref{genrealsol}) and (\ref{qtanalyticform}) in Eq. (\ref{auxiliaryeqns}), we obtain,
\begin{equation}
\label{finalxr}
t-t_s\simeq \frac{2^{3/4} 3^{3/8} f_0^{3/4}}{R^{3/8}} \, .
\end{equation}
Then by using Eqs. (\ref{qtanalyticform}),
(\ref{finalxr}) and (\ref{r1}), we easily obtain the resulting $F(R)$ gravity, which has a particularly interesting form,
\begin{equation}
\label{finalfrgravity}
F(R)\simeq C_1 R+\frac{81 C_1 f_0^6 }{1971200}R^2\, ,
\end{equation}
which is an $R^2$ gravity, with the difference that in the case at hand, the coefficient of the $R^2$ term is very large, in comparison to the $R^2$ inflation model. By using other values of the parameter $\alpha$, the resulting $F(R)$ gravity is different from the one of Eq. (\ref{finalfrgravity}), for example for $a=8/3$, then by repeating the procedure, we obtain the result,
\begin{equation}\label{finalresult1}
F(R)\simeq \left(\frac{ \beta_1}{2 \left(\frac{\beta_3^{3/16}}{\beta_1^{3/16}}\right)^{16/3}}-\frac{ \beta_3}{4 \left(\frac{\beta_3^{3/16}}{\beta_1^{3/16}}\right)^{32/3}}\right)R^2+\frac{ \beta_1^{27/16} \beta_2}{2\ 2^{11/16} \beta_3^{27/16}}R^{27/16}\, ,
\end{equation}
with $\beta_i$ constant parameters, which again provides a leading order behavior of the $F(R)$ gravity. We need to stress that both the $F(R)$ gravity functions of Eq. (\ref{finalfrgravity}) and (\ref{finalresult1}), are actually leading order results and not the exact forms of the $F(R)$ gravities that generate the late-time behavior of the cosmology (\ref{deformedscalefactor}). Therefore, these may originate from $F(R)$ gravity models which pass all the large scale phenomenological tests and also satisfy the local astrophysical constraints, like for example the Hu-Sawicki model \cite{sawicki}, which is described by the $F(R)$ gravity,
\begin{equation}\label{sawickimodel}
F(R)=R-\frac{c_2R^n}{c_1R^n+1}\, ,
\end{equation}
with $n>0$. For example, in the case $n=1$, the Hu-Sawicki model as $R\to 0$, at leading order behaves as follows,
\begin{equation}\label{approx1sawicki}
F(R)\simeq (1-c_2) R+c_1 c_2 R^2\, ,
\end{equation} 
so if we choose the parameters $c_1$ and $c_2$ as follows,
\begin{equation}\label{oneinamil}
C_1=1-c_2,\,\,\,c_1 c_2=\frac{81 C_1 f_0^6 }{1971200}\, ,
\end{equation}
the results of Eq. (\ref{finalresult1}) and (\ref{approx1sawicki}) are identical at leading order. In principle, the same study could be performed for other Types of finite time singularities, like for example the Type II case, in which case by choosing $\alpha=2/3$, the leading order behavior of the $F(R)$ gravity reads,
\begin{equation}\label{leadingorder}
F(R)\simeq 12\,c_2 R+54\ 30^{2/3} f_0^2c_2 R^{2/3} -\frac{24\ 2^{1/3} 3^{5/6} f_0\,c_2}{5^{1/6}}R^{5/6}\, .
\end{equation}

Finally, before closing, let us calculate the EoS for the $F(R)$ gravity, which is now defined to be,
\begin{equation}\label{weff}
w_{eff}=-1-\frac{2\dot{H}}{3H^2}\, ,
\end{equation}
and note the difference between the LQC EoS of Eq. (\ref{eosdef}), however the two definitions coincide when $\rho_c\to \infty$. In the case at hand, since $H(t)\simeq f_0 (t-t_s)^{\alpha}$ near the Rip times, the resulting EoS is approximately equal to,
\begin{equation}\label{deanceintothefire}
w_{eff}\simeq -1-\frac{2 (t_s-t)^{-1+\alpha } \alpha }{3 f_0}\, ,
\end{equation}
and since $\alpha>1$, the EoS is $w_{eff}< -1$, which describes a phantom acceleration era before the Rip is reached. Hence even in the $F(R)$ gravity case, the Universe is accelerating during the last stages of its evolution, before the Rip, with a phantom behavior.

\section{Conclusions}

In this paper we demonstrated that in the context of LQC, it is possible to realize a deformed matter bounce scenario, in which the deformation practically alters the late-time behavior of the model. Specifically, the deformed matter bounce scenario describes a dark energy era, which ends abruptly in a Rip singularity. To be specific, the Rip singularity cannot be reached by the physical system, since the energy density and pressure become complex, nearly a second before the singularity itself is reached. This indicates that this scenario ends before the singularity, and the physical theory must be enriched with some fundamental quantum gravity theory, yet to be found.

We described in detail the qualitative features of the model and we showed that during almost all eras, the model is practically indistinguishable from the standard matter bounce scenario, so all the appealing features of the matter bounce scenario, such as the production of a scale invariant spectrum, are also present in the deformed matter bounce scenario, and of course all the drawbacks, such as an initial non-causal state of the Universe, are also present in the deformed model too. Particularly, the deformed matter bounce model has the following features: at early times and during the matter domination era, the Universe is non-causal with a scale invariant power spectrum and it contracts until a minimal radius is reached, at which point it bounces off and hence the initial singularity of the standard inflationary paradigm is avoided. After the bounce, the Universe expands in a decelerating way, until late-times where the Universe starts to expand in an accelerating way, and continues to expand in this way until a Rip singularity is reached, however before the singularity is reached, the expansion stops. Therefore, a new feature of the deformed matter bounce scenario is that the infinite repetition of the bouncing process ends before the Rip singularity. However, we need to stress the fact that prior to the Rip, bound objects, like galaxies or clusters, will be ripped apart, at a time that depends on the model under study. For example in Ref. \cite{kamio}, by using general relativity arguments the authors have shown that in the case of a phantom dark energy era, our galaxy would be stripped roughly 60 million years before the Rip, and clusters would be stripped a billion years before the Rip, see \cite{kamio}. This could be the case for our deformed model, but we did not address this issue in this paper, since we focused more on the fact that the deformed matter bounce model is a ``one way model'' in the sense that the repetition of the bouncing process stops before the Rip, regardless what happens before the Rip. This is clearly different from the standard LQC matter bounce scenario, in which the Universe infinitely oscillates in a matter dominated way.

After giving a detailed qualitative description of the model, we investigated how the deformed matter bounce cosmology can be described by a viscous imperfect fluid, in the context of LQC and as we demonstrated, in the LQC case, the resulting EoS is of the form $P=-\rho-B(H,\dot{H})$. In addition to this study, we examined how the deformed matter bounce scenario can be realized in the case that the matter fluid corresponds to a canonical scalar field, always in the context of LQC. The resulting picture is quite interesting, since during the matter domination era, which occurs during the contraction and the expansion era, the potential is identical to the one found in the literature, while at late times, the potential is of power law form. 

We need to note that in the present work we generalized the matter bounce cosmological scenario, but it is possible to use more general scenarios, like for example the radiation bounce \cite{lcdmcai,lcdmsergei}, in which case the EoS is $p=\frac{1}{3}\rho$, or for example bouncing models with EoS $p=w\rho$, but we did not go into details, because the procedure is the same. Finally, by repeating the line of research we adopted in this paper, we can achieve the transition from the bounce to the dark energy epoch ending in some soft singularity of Type II, III and IV \cite{Nojiri:2005sx}.

\section*{Acknowledgments}

This work is supported by MINECO (Spain), project
 FIS2013-44881 (S.D.O) and by Min. of Education and Science of Russia (S.D.O
and V.K.O).

\end{document}